\documentclass[]{emulateapj}
\usepackage{amsmath}
\usepackage{natbib}
\usepackage{color}
\usepackage{graphicx}
\graphicspath{ {images/} }
\usepackage{url}
\newcommand{\hst}{{\it HST}}

\newcommand{\Planck}{{\it Planck}}
\newcommand{\lenstool}{\texttt{Lenstool}}

\shorttitle{Strong Gravitational Lens Models for RELICS Clusters}
\slugcomment{Accepted to ApJ: 4/13/18}
\shortauthors{Cerny et al.}

\begin{document}
\title{RELICS: Strong Lens Models for Five Galaxy Clusters From the Reionization Lensing Cluster Survey}

\author{Catherine Cerny\altaffilmark{1},
Keren Sharon\altaffilmark{1},
Felipe Andrade-Santos\altaffilmark{2},
Roberto J. Avila\altaffilmark{3},
Maru\v{s}a Brada\v{c}\altaffilmark{4},
Larry D. Bradley\altaffilmark{3},
Daniela Carrasco\altaffilmark{6},
Dan Coe\altaffilmark{3},
Nicole G. Czakon\altaffilmark{7},
William A. Dawson\altaffilmark{8},
Brenda L. Frye\altaffilmark{9},
Austin Hoag\altaffilmark{4},
Kuang-Han Huang\altaffilmark{4},
Traci L. Johnson\altaffilmark{1},
Christine Jones\altaffilmark{2},
Daniel Lam\altaffilmark{5},
Lorenzo Lovisari\altaffilmark{2},
Ramesh Mainali\altaffilmark{9},
Pascal A. Oesch\altaffilmark{10},
Sara Ogaz\altaffilmark{3},
Matthew Past\altaffilmark{1},
Rachel Paterno-Mahler\altaffilmark{1,}\altaffilmark{14},
Avery Peterson\altaffilmark{1},
Adam G. Riess\altaffilmark{3},
Steven A. Rodney\altaffilmark{11},
Russell E. Ryan\altaffilmark{3},
Brett Salmon\altaffilmark{3},
Irene Sendra-Server\altaffilmark{12,}\altaffilmark{15},
Daniel P. Stark\altaffilmark{9},
Louis-Gregory Strolger\altaffilmark{3},
Michele Trenti\altaffilmark{6},
Keiichi Umetsu\altaffilmark{7},
Benedetta Vulcani\altaffilmark{6},
Adi Zitrin\altaffilmark{13}
}

\email{krosah@umich.edu}
\altaffiltext{1}{Department of Astronomy, University of Michigan, 1085 South University Ave, Ann Arbor, MI 48109, USA}
\altaffiltext{2}{Harvard-Smithsonian Center for Astrophysics, 60 Garden Street, Cambridge, MA 02138, USA}
\altaffiltext{3}{Space Telescope Science Institute, 3700 San Martin Drive, Baltimore,
MD 21218, USA}
\altaffiltext{4}{Department of Physics, University of California, Davis, CA 95616, USA}
\altaffiltext{5}{Leiden Observatory, Leiden University, NL-2300 RA Leiden, The Netherlands}
\altaffiltext{6}{School of Physics, University of Melbourne, VIC 3010, Australia} 
\altaffiltext{7}{Institute of Astronomy and Astrophysics, Academia Sinica, PO Box 23-141, Taipei 10617,Taiwan}
\altaffiltext{8}{Lawrence Livermore National Laboratory, P.O. Box 808 L-210, Livermore, CA, 94551, USA}
\altaffiltext{9}{Department of Astronomy, Steward Observatory, University of Arizona, 933 North Cherry Avenue, Rm N204, Tucson, AZ, 85721, USA}
\altaffiltext{10}{Department of Astronomy, Yale University, New Haven, CT 06520, USA}
\altaffiltext{11}{Department of Physics and Astronomy, University of South Carolina, 712 Main St., Columbia, SC 29208, USA}
\altaffiltext{12}{American School of Warsaw, Warszawska 202, 05-520 Bielawa, Poland}
\altaffiltext{13}{Physics Department, Ben-Gurion University of the Negev, P.O. Box 653, Beer-Sheva 84105, Israel}
\altaffiltext{14}{Department of Physics and Astronomy, University of California, Irvine, CA 92697, USA}
\altaffiltext{15}{Department of Theoretical Physics, University of Basque Country UPV/EHU, E-48080 Bilbao, Spain}

\begin{abstract}
Strong gravitational lensing by galaxy clusters magnifies background galaxies, enhancing our ability to discover statistically significant samples of galaxies at $z>6$, in order to constrain the high-redshift galaxy luminosity functions.  
Here, we present the first five lens models out of the Reionization Lensing Cluster Survey (RELICS)  Hubble Treasury Program, based on new \hst\ WFC3/IR and ACS imaging of the clusters RXC~J0142.9$+$4438, Abell~2537, Abell~2163, RXC~J2211.7$-$0349, and ACT$-$CLJ0102$-$49151. The derived lensing magnification is essential for estimating the intrinsic properties of high-redshift galaxy candidates, and properly accounting for the survey volume.
We report on new spectroscopic redshifts of multiply imaged lensed galaxies behind these clusters, which are used as constraints, and detail our strategy to reduce systematic uncertainties due to lack of spectroscopic information. In addition, we quantify the uncertainty on the lensing magnification due to statistical and systematic errors related to the lens modeling process, and find that in all but one cluster, the magnification is constrained to better than 20\% in at least 80\% of the field of view, including statistical and systematic uncertainties.
The five clusters presented in this paper span the range of masses and redshifts of the clusters in the RELICS program. We find that they exhibit similar strong lensing efficiencies to the clusters targeted by the Hubble Frontier Fields within the WFC3/IR field of view.
Outputs of the lens models are made available to the community through the Mikulski Archive for Space Telescopes.

\end{abstract}

\keywords{galaxies: clusters: individual (RXCJ0142.9+4438, Abell2537, Abell2163, RXCJ2211.7-0349, ACT-CLJ0102-49151) --- gravitational lensing: strong }

\section{Introduction}
\begin{deluxetable*}{lccccccc}[!bp]
\tablecolumns{6}
\tablecaption{Cluster Properties}
\tablehead{\colhead{Cluster }  &
            \colhead{R.A. (J2000)}    &
            \colhead{Dec (J2000)}    &
            \colhead{$z$}    &
            \colhead{Mass ($10^{14}M_\sun$)}    &} 
\startdata
Abell~2537            & 23:08:21  & $-$02:11:44   & 0.2966  & 5.52	   \\
RXC~J0142.9+4438    & 01:42:53   & 44:38:20	 & 0.3410  & 9.02           \\
RXC~J2211.7-0349     & 22:11:43  & $-$03:49:45   & 0.3970  & 10.50 \\
ACT-CLJ0102-49151   & 01:02:56   & $-$49:16:18  & 0.8700  & 10.75 \\
Abell~2163            & 16:15:49  & $-$06:09:08   & 0.2030  & 16.12 
\enddata
\tablecomments{Clusters considered in this work. The mass estimate is $M_{500}$ from \Planck\ Collaboration (2015).}
\label{tab.clusters}
\end{deluxetable*}

The search for high-redshift galaxies is central to the study of early galaxy formation and evolution. The galaxy luminosity function can be used as a tool to determine the distribution and statistical properties of galaxies throughout the universe over cosmic time, and can lend insight into the early stages of galaxy evolution (Livermore et al. 2017). However, constraining this function is challenging at redshifts greater than $z\sim9$ (Bouwens et al. 2015). This time period marks the epoch of reionization (\Planck\ Collaboration 2016, Robertson et al. 2015), which is not yet well understood {due} to the small number of sources that have been detected at such high redshifts. 
{Without lensing magnification, most observable high-redshift galaxies are drawn from the luminous end of the luminosity function as a result of cosmological dimming. These galaxies are thus not representative of their population.} 

The Reionization Lensing Cluster Survey (RELICS; Coe et al., in preparation) seeks to discover a statistically significant sample of galaxies at high redshifts, in order to better constrain the luminosity functions and improve subsequent study of the epoch of reionization. It aims to achieve this scientific goal by combining the high resolution and infrared capabilities of \hst\, with the magnification boost of strong gravitational lensing by galaxy clusters. 
Contrary to extremely deep lensing programs such as the Hubble Frontier Fields (Lotz et al. 2017), RELICS maximizes the probability of discovering galaxies at high redshift by conducting  shallow observations of a large area.
Our 190-orbit \hst\ Treasury Program (GO 14096; PI Coe) observed 41 clusters with ACS and WFC3/IR, providing the first \hst\ infrared images of these fields. Additionally, 390 hours of Spitzer imaging (PI Brada\v{c}, PI Soifer) support the high redshift search and help constrain galaxy properties.

{The lensing potential of each cluster must be quantified to fully exploit the strong lensing boost of the foreground galaxy clusters. This potential is measured by computing a detailed lens model that determines the projected mass density distribution of the lensing cluster.} The derived lensing magnification is required for converting observed measurements such as size, luminosity, star formation rate, and stellar mass of lensed galaxies to their intrinsic physical properties. Additionally, the lens models provide a magnification correction for the survey volume, which is used to normalize the luminosity functions.
RELICS will publish lens models for each strong lensing cluster, based on the observed locations of strongly lensed, multiply imaged, background galaxies. The redshifts of these lensed galaxies {are} especially important for the accuracy of the models, as inaccurate redshifts can significantly bias the derived magnification (Johnson \& Sharon, 2016). We employ ground-based spectroscopic observations in conjunction with \hst\ photometry in order to better constrain our lens models.

The lens modeling community has taken an increasingly serious look at systematic biases in recent years. In this paper, we continue this effort by investigating the systematic uncertainties due to the lack of spectroscopic redshifts of lensed galaxies. We detail our new strategy for reducing lens modeling bias {through} a careful incorporation of the photometric redshift posterior into the lens modeling process. 

Here, we present strong lens models for five clusters from the first set of RELICS Cycle~24 observations. We describe the sample selection in Section~2,  and the \hst\ imaging and ground-based spectroscopy followup in Section~3. We then describe the lens modeling process we used and show the lens models and best-fit parameters in Section~4. We assess the lens modeling uncertainties in Section~5. 
Finally, in Section~6, we discuss our results and consider how our methods can be applied to future work in this survey and beyond.
We assume \( \Lambda \)CDM cosmology throughout this paper with \( \Omega_M \)=0.3, \(\Omega_{\Lambda} \)=0.7, and \(H_0\) =70 km s\(^{-1}\) Mpc\(^{-1}\). All magnitudes are measured in the AB system. 
We use the standard notation $M_\Delta$ to denote the mass enclosed
within a sphere of $r_\Delta$, within which the mean overdensity
equals $\Delta\times \rho_\mathrm{c}$, where $\rho_\mathrm{c}$ is the
critical density of the universe at the cluster redshift. All images are oriented North-East, with North up and East to the left.

\begin{deluxetable*}{llcccccc}[!hbtp]
\tablecolumns{6}
\tablecaption{Cluster Imaging Parameters}
\tablehead{\colhead{Cluster }  &
            \colhead{Filter}    &
            \colhead{Date of Observation}       &
            \colhead{Exposure time [s]} &
            \colhead{PID}
                                    }
\startdata 
Abell~2537  &  ACS F435W       & 2016-06-08 &        2162    & RELICS \\ 
            &  ACS F606W       & 2002-10-02 &       2080     & GO9270 \\ 
            &  ACS F814W       & 2016-07-19 &       2162     & RELICS \\ 
            &  WFC3/IR F105W   & 2016-07-19 &        756     & RELICS \\ 
            &                  & 2016-08-06 &        756     & RELICS \\ 
            &  WFC3/IR F125W   & 2016-07-19 &        356     & RELICS \\ 
            &                  & 2016-08-06 &        356     & RELICS \\ 
            &  WFC3/IR F140W   & 2016-07-19 &        356     & RELICS \\ 
            &                  & 2016-08-06 &        356     & RELICS \\ 
            &  WFC3/IR F160W   & 2016-07-19 &        1006    & RELICS \\ 
            &                  & 2016-08-06 &        956     & RELICS \\ 
\hline
RXC~J0142.9+4438    & ACS F435W         & 2015-12-04   &    2268    & RELICS \\
                    & ACS F606W         & 2016-01-14   &    2189    & RELICS \\
                    & ACS F814W         & 2015-12-04   &    2439    & RELICS \\
                    & WFC3/IR F105W     & 2015-12-04   &    756     & RELICS  \\
                    &                   & 2016-01-14   &    756     & RELICS  \\
                    & WFC3/IR F125W     & 2015-12-04   &    356     & RELICS  \\
                    &                   & 2016-01-14   &    406     & RELICS  \\
                    & WFC3/IR F140W     & 2015-12-04   &    381     & RELICS  \\
                    &                   & 2016-01-14   &    406     & RELICS  \\
                    & WFC3/IR F160W     & 2015-12-04   &    1106    & RELICS  \\
                    &                   & 2016-01-14   &    956     & RELICS  \\
\hline
RXC~J2211.7-0349    & ACS F435W         & 2016-11-25   & 1953       & RELICS \\
                    & ACS F606W         & 2011-11-19   & 1200       & GO12166 \\
                    &                   & 2015-11-25   & 2101       & RELICS \\
                    & ACS F814W         & 2016-11-25   & 2124       & RELICS \\
                    & WFC3/IR F105W     & 2015-10-16   &    706     & RELICS  \\
                    &                   & 2015-11-25   &    706     & RELICS  \\
                    & WFC3/IR F125W     & 2015-10-16   &    356     & RELICS  \\
                    &                   & 2015-11-25   &    356     & RELICS  \\
                    & WFC3/IR F140W     & 2015-10-16   &    331     & RELICS  \\
                    &                   & 2015-11-25   &    356     & RELICS  \\
                    & WFC3/IR F160W     & 2015-10-16   &    906     & RELICS  \\
                    &                   & 2015-11-25   &    1006    & RELICS  \\
\hline
ACT-CLJ0102-49151   & ACS F435W         & 2016-07-08   & 2093       & RELICS \\
                    & ACS F606W         & 2012-12-22   & 1920       & GO12477 \\
                    & WFC3/IR F105W     & 2016-07-09   &    706     & RELICS  \\
                    &                   & 2016-08-08   &    756     & RELICS  \\
                    & WFC3/IR F125W     & 2016-07-09   &    356     & RELICS  \\
                    &                   & 2016-08-08   &    381     & RELICS  \\
                    & WFC3/IR F140W     & 2016-07-09   &    356     & RELICS  \\
                    &                   & 2016-08-08   &    381     & RELICS  \\
                    & WFC3/IR F160W     & 2016-07-09   &    1006    & RELICS  \\
                    &                   & 2016-08-08   &    1006    & RELICS  \\
\hline
Abell~2163  & ACS F435W         & 2011-07-03   &      4664    & GO12253 \\
            & ACS F606W         & 2011-07-03   &      4667    & GO12253 \\
            & ACS F814W         & 2011-07-03   &    9192      & GO12253 \\
            & WFC3/IR F105W     & 2016-09-03   &    706       & RELICS  \\
            &                   & 2016-04-18   &    706       & RELICS  \\
            & WFC3/IR F125W     & 2016-09-03   &    356       & RELICS  \\
            &                   & 2016-04-18   &    356       & RELICS  \\
            & WFC3/IR F140W     & 2016-09-03   &    356       & RELICS  \\
            &                   & 2016-04-18   &    356       & RELICS  \\
            & WFC3/IR F160W     & 2016-09-03   &    1006      & RELICS  \\
            &                   & 2016-04-18   &    1006      & RELICS  

\enddata
\tablecomments{Dates, filters, and exposure times for each cluster, including both RELICS observations and archival data from other proposals.
} \label{tab.hstobs}
\end{deluxetable*}

\section{Sample Selection}
The RELICS program observed a total of 46 fields lensed by 41 galaxy clusters, {which were selected} based on their Sunyaev Zel'dovich effect (SZ) or Xray-inferred mass and other criteria. The sample will be described in full detail in a forthcoming paper (Coe et al., in prep). 

About half of the RELICS clusters were mass-selected from the \Planck\ SZ cluster catalog (\Planck\ Collaboration 2016). The \Planck\ cluster catalog provides redshifts and mass estimates for 1094 clusters. Of the 34 most massive clusters ($M_{500} > 8.8 \times 10^{14} M_{\odot}$), only 13 have already been imaged by \hst\ in the IR prior to Cycle~23. The remaining 21 clusters form our mass-selected sample. 
These clusters are expected to have X-ray masses similar to or
greater than clusters in the Frontier Fields program (Lotz et al. 2017) and the CLASH program (Postman et al. 2012); the average mass in CLASH is 
$M_{500} \simeq 1.0 \times 10^{15} M_{\odot} h^{-1}_{70}$ 
(Umetsu et al. 2014, 2016), and the lowest 
$M_{500}$ mass is $4-5 \times 10^{14} M_{\odot} h^{-1}_{70}$.
High mass clusters are likely to have a large cross-section for strong lensing, making them ideal candidates for the survey.

To increase the efficiency of the program, the other 20 clusters were selected {from} clusters that have previously been identified as prominent strong lenses based on available imaging. The selection of these 20 clusters also relied on X-ray mass estimates (MCXC compilation Piffaretti et al. 2011; Mantz et al. 2010); weak lensing mass estimates (Sereno et. al 2015 compilation, including Weighing the Giants, Applegate et al. 2014, von der Linden et al. 2014; Umetsu et al. 2014; Hoekstra et al. 2015); and SZ mass estimates from SPT (Bleem et al. 2015) and ACT (Hasselfield et al. 2013). We also considered a range of clusters from the SDSS survey (Wong et al. 2013, Wen et al. 2012) and clusters nearly selected for the Frontier Fields.
The survey design thus maximizes the chances of finding high redshift galaxies.

The clusters presented in this paper (Table~\ref{tab.clusters}) are the first five clusters with completed models. Lens models for the remaining 36 cluster fields will be presented in future papers.

 \section{Observations}
\subsection{\hst\ Imaging}
We obtained optical and near-infrared HST photometric data of 41 galaxy cluster fields with the Wide Field Camera 3 (WFC3) and the Advanced Camera for Surveys (ACS). We use four WFC3/IR filters and three ACS filters, which span {a} wavelength range of $0.4-1.7$ microns. These filters are used over a total of 190 orbits to image 46 fields lensed by 41 clusters.\footnote{20 of the orbits are designated as ToO to follow-up lensed supernova candidates.} Each cluster was observed for a total of five orbits, except for cases where archival data from the ACS {were} available. In these cases, the number of orbits was reduced accordingly. Each cluster was observed in two epochs separated by $40-60$ days in order to identify supernovae and other transient phenomena. Table~\ref{tab.hstobs} lists the observing dates and exposure times for RELICS observations of the fields considered in this work, and provides the proposal identification information for archival data.

\begin{deluxetable}{lcl}
\tablecolumns{3}
\tablecaption{Cluster Spectroscopy Observations}
\tablehead{\colhead{Cluster }  &
            \colhead{Date}    &
            \colhead{Exposure Time, Notes}}
\startdata
Abell~2163 &  2016 Jun 07-08      & 1h \\
Abell~2537        & 2016 Aug 02-03      & 2 masks, 1.5h each \\
RXC~J2211.7-0349 & 2016 Aug 02-03      & 2h \\
ACT-CLJ0102-49151 &  2016 Aug 02-03      & 2h 
\enddata
\tablecomments{Dates and exposure times for spectroscopic observations carried out with Magellan/LDSS3 on the clusters presented in this paper to date. {The average seeing during integration for each cluster was $0.7\arcsec$, with the exception of ACT-CLJ0102, where the average seeing was $0.8\arcsec$.}} 
\label{tab.specobs}
\end{deluxetable}

\subsection{Imaging Data Reduction}
All sub-exposures in each filter were combined to form a deep image using the AstroDrizzle package (Gonzaga et al. 2012) using \texttt{PIXFRAC}=0.8. The data in different filters were aligned to the same pixel frame, and the astrometry was corrected to match the Wide-field Infrared Survey Explorer (WISE) point source catalog (Wright et al. 2010). Final drizzled images were generated in two pixel scales: $0\farcs03$ and $0\farcs06$, for best sampling of the ACS and WFC3/IR point spread functions, respectively. The ACS data were corrected for charge transfer inefficiency losses prior to drizzling.
The final reduced data are made available to the public as high level science products through the Mikulski Archive for Space Telescopes (MAST).\footnote{https://archive.stsci.edu/prepds/relics/}

\subsection{Photometry Catalog}
The RELICS team {has produced} catalogs of photometry and photometric redshifts for each field, based on the final $0\farcs06$ dataset. {These catalogs are} made available through MAST. A full description of the photometry data products will appear in a separate publication (Coe et al., in prep).
Sources are extracted from {a} weighted stack of all the data (ACS and WFC3/IR), and from {a weighted stack of the} WFC3/IR data alone using SExtractor (Bertin \& Arnouts 1996) version 2.8.6 in dual-image mode. 
The weighted stack image is used as the reference image in SExtractor for all seven filters.
The fluxes and colors {of sources} are measured within isophotal apertures, and are corrected for Galactic extinction (Schlafly \& Finkbeiner 2011). 

The photometric redshift (photo-z) of each galaxy is estimated using the BPZ algorithm (Bayesian Photometric Redshifts; Benitez et al. 2000, 2014; Coe et al. 2006). 
We run BPZ with 11 spectral energy distribution (SED) model templates (five ellipticals, two late types, and four starbursts) as described in Benitez et al. (2014) and Rafelski et al. (2015).  The models are based on PEGASE including emission lines (Fioc \& Rocca-Volmerange 1997). {They are} then recalibrated to match the observed photometry of galaxies with spectroscopic redshifts from FIREWORKS (Wuyts et al. 2008). {The BPZ algorithm} redshifts these SEDs by fitting them to the observed photometry of each galaxy, using a Bayesian prior on redshift given the $i$-band magnitude and spectral type. The elliptical templates for galaxies at high redshifts are downweighted by the prior, and lower redshift solutions are generally preferred over those at higher redshifts. Possible lensing magnification is not considered in the prior.

Strongly lensed galaxies, in particular those used as constraints in this paper, were {given further manual inspection. Several of these} sources yielded less reliable photometric redshifts because they were fainter and/or contaminated by brighter nearby galaxies (usually cluster members). {As a result,} the photometric redshifts of multiple lensed images match one another often, but not always. 

The system photo-z was manually determined by examining the redshift solution of all the images of the same system, {where higher weight was given} to images that are brighter and better isolated. We excluded any redshift solutions lower than the cluster redshift for {systems that were identified as lensed sources to ensure that the model placed the source behind the cluster.}

\subsection{Ground-Based Spectroscopy}\label{s.spectroscopy}

Ground-based spectroscopic observations of RELICS clusters were conducted using the upgraded Low Dispersion Survey Spectrograph (LDSS3-C\footnote{http://www.lco.cl/telescopes-information/magellan/ instruments/ldss-3}) on the Magellan Clay telescope using {time from the} University of Michigan allocation (PI: Sharon) and the University of Arizona allocation (PI: Stark). {The observations used} multi-object slit masks, {which had} slits placed on candidate lensed galaxies at highest priority. The remaining slits were placed on cluster member galaxies (selected by color) or other
interesting objects in the cluster field, such as candidate ram-pressure stripping galaxies (e.g., Ebeling et al. 2014) or candidate supernova hosts. We used the VPH-ALL grism, which has high sensitivity over the widest range of wavelengths: 4250 \AA $<\lambda<$10000 \AA. {This range is} useful for the identification of emission lines at unknown redshifts.
A $1\farcs0$ slit was used for all objects, which yielded a spectral resolution of R~450-1100 across that wavelength range\footnote{http://www.lco.cl/telescopes-information/magellan/ instruments/specs/LDSS-3\%20Handout.pdf}. The detector covers a spatial extent of $6.4$ arcmin. 

Table~\ref{tab.specobs} provides the date of the observation and the exposure time for each field considered in this work.
The spectroscopic data were reduced using the standard procedures using the COSMOS data reduction package.\footnote{http://code.obs.carnegiescience.edu/cosmos} We report here on spectroscopic redshifts that were secured for candidate lensed galaxies and used as constraints in this paper. A full description of the results from these follow-up programs will be given in a future publication {(Mainali et al., in prep).}

\section{Strong Lensing Analysis}\label{s.lensing}
We present the lens models for five galaxy clusters in the RELICS survey. The properties of these clusters are given in Table~\ref{tab.clusters}.

Strong lens models were computed using \lenstool\ (Jullo et al. 2007), which models the cluster projected mass density as a combination of parametric halos. Each halo is modeled as a pseudo-isothermal ellipsoidal mass distribution (PIEMD; Limousin et al. 2005) and the best-fit set of parameters is found using Markov Chain Monte Carlo (MCMC) sampling. The parameters that are allowed to vary in the modeling process are, for each halo, the $x$ and $y$ position of the center of the halo; the ellipticity of the halo, $e$; its position angle, $\theta$; the core radius, $r_{core}$; {the truncation radius of the cluster halo, $r_{cut}$}; and the effective velocity dispersion, $\sigma$. Optimization of the models presented in this paper was performed in the image plane.

Cluster-member galaxies were selected using the cluster red sequence method (Gladders \& Yee, 2000) {to have colors consistent with the red sequence at this redshift. Two bands that straddle the 4000\AA $ $  break were used for each cluster depending on redshift (F606W-F814W for Abell~2163 and RXC~J2211.7-0349; F814W-F105W for Abell~2537; F435W-F814W for RXC~J0142.9+4438; F105W-F140W for ACT-CLJ0102-49151). The brightest 100 galaxies were used for Abell~2163, Abell~2537, and ACT-CLJ0102-49151, while the brightest 166 galaxies were used for RXC~J2211.7-0349 and the brightest 200 galaxies were used for RXC~J0142.9+4438.}

The selected cluster galaxies were modeled as PIEMD halos. Following Limousin et al. (2005), the core radius of the cluster member galaxies was fixed at 0.15 kpc, and $r_{cut}$ and  $\sigma$ were scaled with the F814W luminosity, except for ACT-CLJ0102-49151, for which the F105W band was used. The positional parameters ($x$, $y$, $e$, $\theta$) were fixed to the properties of their light distribution as measured with SExtractor (Bertin et al. 1996).

{The strong lens models were constrained by identifying the positions of multiply-imaged systems (``arcs'').} Arcs are identified by eye as multiple images of the same background galaxy, based on their morphology, structure, and color. When we identify a system of arcs, we verify that the orientation and parity of each arc is consistent with the expectation from lensing. {We assume an uncertainty of $0.3\arcsec$ in the positions of each arc.}

{Several authors (e.g., Smith et al. 2009, Johnson \& Sharon 2016) have shown that even with excellent positional constraints, such as the ones provided by \hst\ imaging, the accuracy of strong lens models relies on the availability of redshift information of the lensed galaxies. In particular, models that lack redshift constraints are limited in their ability to correctly measure the slope of the mass distribution, and the lensing magnification derived from such models may be biased (Johnson \& Sharon 2016). 

The RELICS program aims to provide the community with the best available lens models of {all observed clusters in time to prepare} for JWST observations. However, {as of the release of this paper} some clusters in the survey have no redshift constraints, some clusters have a spectroscopic redshift of only one lensed galaxy, and some have two or more spectroscopic redshifts. {When spectroscopic information is unavailable, photometric redshifts can be used as constraints} by setting limits on the redshift parameter. These limits are based on the probability distribution function of the photometric redshift analysis. However, catastrophic outliers might bias the result and force the model into a wrong solution, as shown in Johnson \& Sharon (2016). {Thus} it is advised that lens models that are not based on spectroscopic redshifts should be treated with caution.}

To facilitate the release of lens models for all targeted clusters, including those with limited redshift constraints, we follow the procedure below when computing the lens models. The procedure aims to reduce the systematic error due to redshift uncertainty; to reduce the probability of a catastrophic photo-z outlier affecting the result; and to estimate the effect of the redshift uncertainty on deliverables, such as lensing magnification and mass.

{Case~1 -- clusters with at least one spectroscopic redshift of multiply-imaged lensed background sources (Abell~2537, RXC~2211): in this case, we use the spectroscopic redshift(s) as fixed constraint(s).}
The redshifts of arcs without spectroscopic redshifts are treated as free parameters. We use the photometric redshifts of these arcs as a guide to set the redshift prior on the lens model. However, following the recommendations of Johnson \& Sharon (2016), we allow broader limits than the photometric redshift probability distribution function so as not to be biased by catastrophic outliers. 

{Case~2 -- clusters with no spectroscopic redshifts of background sources (RXC~0142, ACT~0102, Abell~2163): in this case, we attempt to leave all the redshifts as free parameters. If this approach fails (e.g., results in unrealistic model-predicted redshifts or lensing mass), we fix the redshift of one of the lensed systems to its best-fit photo-z. {We test the choice of which source to fix to ensure that it has a minor effect} on other model choices. The redshifts of the remaining sources are left as free parameters, as in Case~1.}

{In both cases, we test the agreement between the model-predicted redshifts (model-z) and the photo-z posterior. For a viable lens model, we require that there must not be a significant systematic bias between the model-z and the photo-z, i.e., that the agreement between model-z and photo-z is within the $1\sigma$ uncertainties for most sources. {Our procedure avoids introducing biases into the lens model due to photo-z modeling uncertainties or catastrophic outliers.} In Section~\ref{s.uncertainties}, we quantify the effect of the redshift parameter on the uncertainties. }

In the following sections, we describe each model in more detail. The results of the lens models, {which include the mass and the effective Einstein radius}, are given in Table~\ref{tab.results}. In all the models, our numbering scheme of multiply-imaged lensed sources {lists} the source number first, to the left of the decimal point, and the image number to the right of the decimal point. For example, three images of source galaxy \#1 would be labeled 1.1, 1.2, and 1.3. 
Tables~\ref{tab.A2537arcs}--\ref{tab.2163model} give the details of the lensed galaxies that were used as constraints and the parameters for each lens model, as well as the ID, coordinates, and photo-z of each image. Spectroscopic redshifts are given for sources for which those were measured. {When one source is composed of multiple visually distinct clumps, or emission knots,} photometric redshifts are listed separately for each individual knot.

\subsection{Abell~2537}
\begin{figure*}
    \centering%
    \begin{minipage}{1.0\textwidth}
        \includegraphics[width=0.45\linewidth]{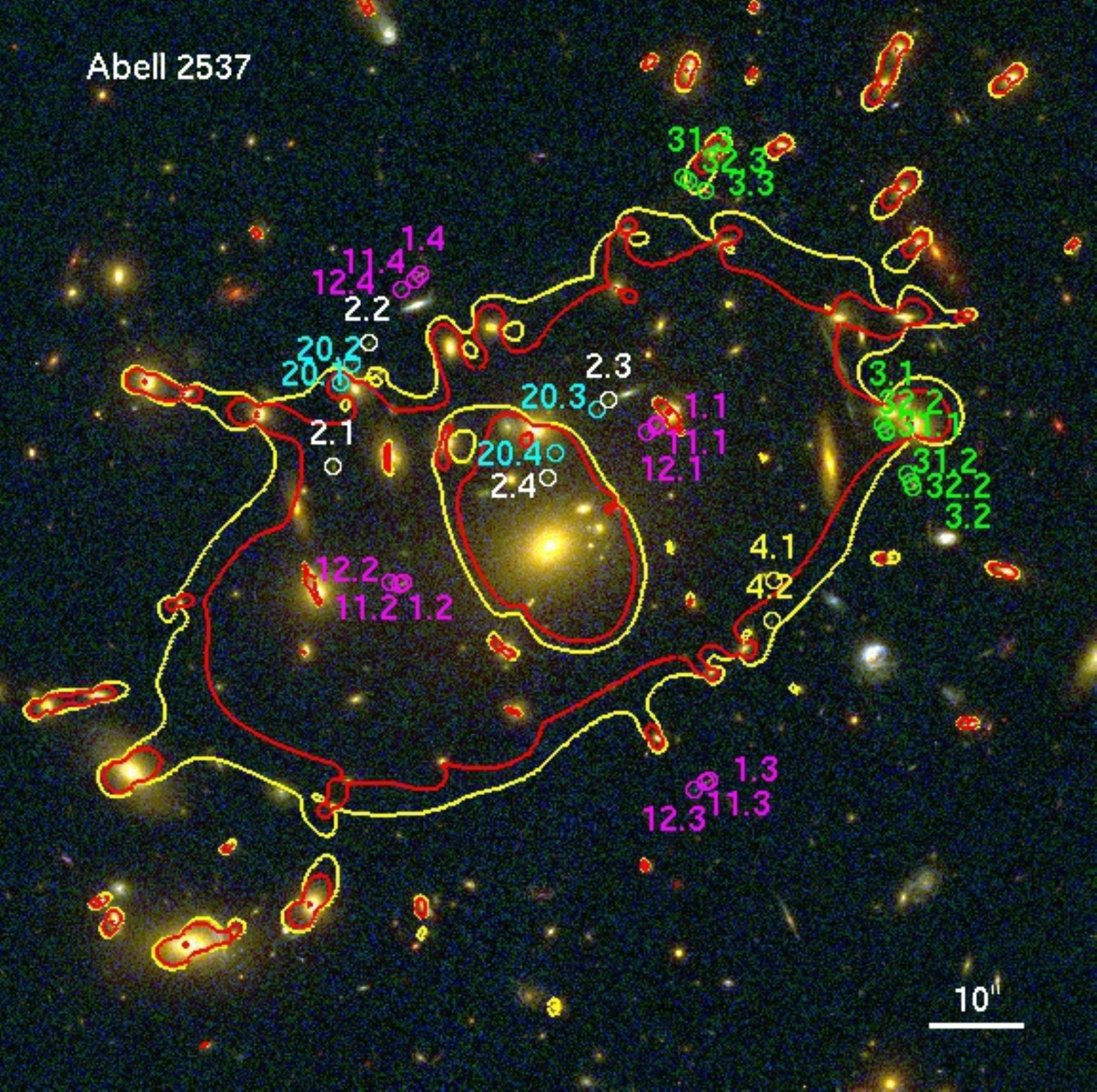}
         \hspace*{0.5cm}
        \includegraphics[width=0.49\linewidth]{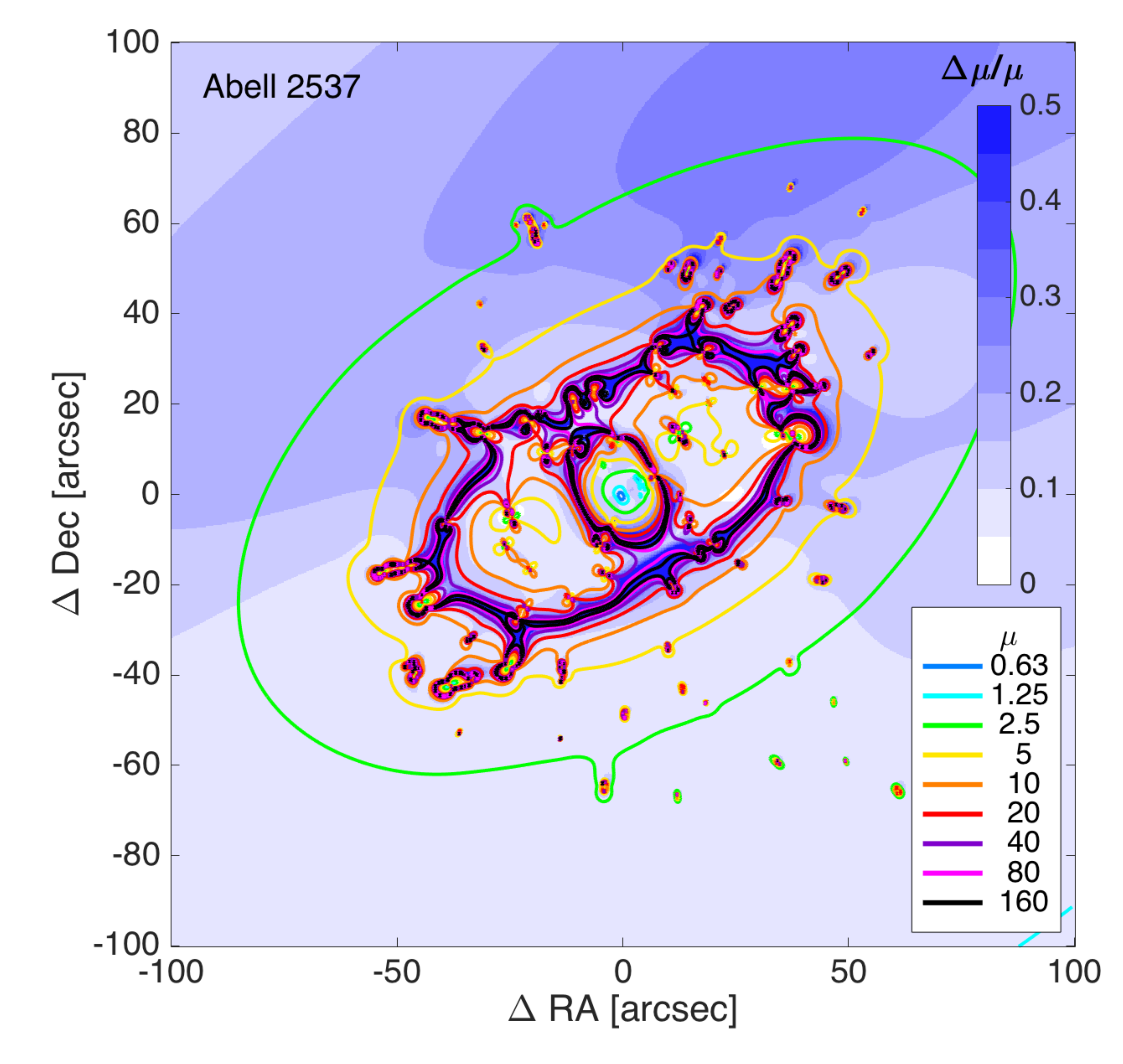}
        \caption{{\it Left}: Image of Abell~2537 created using a combination of WFC3IR imaging in the red (F160W) and ACS imaging in the green (F814W) and blue (F435W). Multiply imaged galaxies are labeled. The red curve marks the location of the critical curve for a source at $z=3.0$. A second critical curve is plotted in yellow at $z=9.0$. The image is oriented North-East.
        {\it Right}: The magnification for a source at $z=9.0$ from the best-fit lens model is shown as contours; contour values are given in the legend. 
        The background colormap indicates the uncertainty level in in each location, given as $\Delta\mu/\mu$. The uncertainty estimate takes into account the source redshift uncertainty, but does not account for other systematics (see text).
        }\label{fig.a2537}
    \end{minipage}%
\end{figure*}

We detect four lensed systems in the field of Abell~2537, as described below (see also Table~\ref{tab.A2537arcs}). 

System \#1 is lensed into four images. We identify three separate emission knots in this galaxy, designated as 1.1-1.4, 11.1-11.4, 12.1-12.4 in Figure~\ref{fig.a2537}. 

System \#2 is lensed into four multiple images, each one containing two emission knots. The images are labeled 2.1-2.4, 20.1-20.4 in Figure~\ref{fig.a2537}. We measure a spectroscopic redshift  of $z_{spec}=3.611$ for image 2.4, based on Ly$\alpha$ emission in our LDSS3 spectroscopy (see Section~\ref{s.spectroscopy}). This falls within the 95\% CL range of the photometric redshifts of the image and is slightly lower than its highest probability peak (Table~\ref{tab.A2537arcs}). We used the spectroscopic redshift as a fixed constraint in our lensing analysis.

System \#3 has three multiple images. We identify three distinct emission knots in each image, labeled 3.1-3.3, 30.1-30.3, 31.1-31.3 in Figure~\ref{fig.a2537}. We find that not fixing a second redshift parameter in this cluster (in addition to the one spectroscopic redshift) causes the model to default to a non-physical result, {e.g. the model predicts redshifts higher than $z=13$}. We therefore fix the redshift of images 3.1-3.3 to the photometric redshift for this source, $z=3.2$, in order to better constrain the lensing potential in the regions further away from the center of the cluster. 

System \#4 is a faint galaxy, labeled 4.1-4.2 in Figure~\ref{fig.a2537}. The lensing analysis predicts additional images at two other locations within the field, but their predicted magnitudes are too faint to be detected, which explains their absence from the data. 

The lens model for this cluster favors a combination of two cluster-scale halos, both located close to the center of the cluster within the strong lensing regime. As described above, we include perturbation from galaxy scale halos. We allow the parameters of the cluster-scale halos to be solved for by the lens model.
Table~\ref{tab.A2537arcs} lists the locations of the lensed galaxies and their redshifts. Table~\ref{tab.A2537model} lists the best-fit parameters of the resulting model.

\subsection{RXC~J0142.9+4438}
\begin{figure*}
    \centering%
    \begin{minipage}{1.0\textwidth}
        \includegraphics[width=0.45\linewidth]{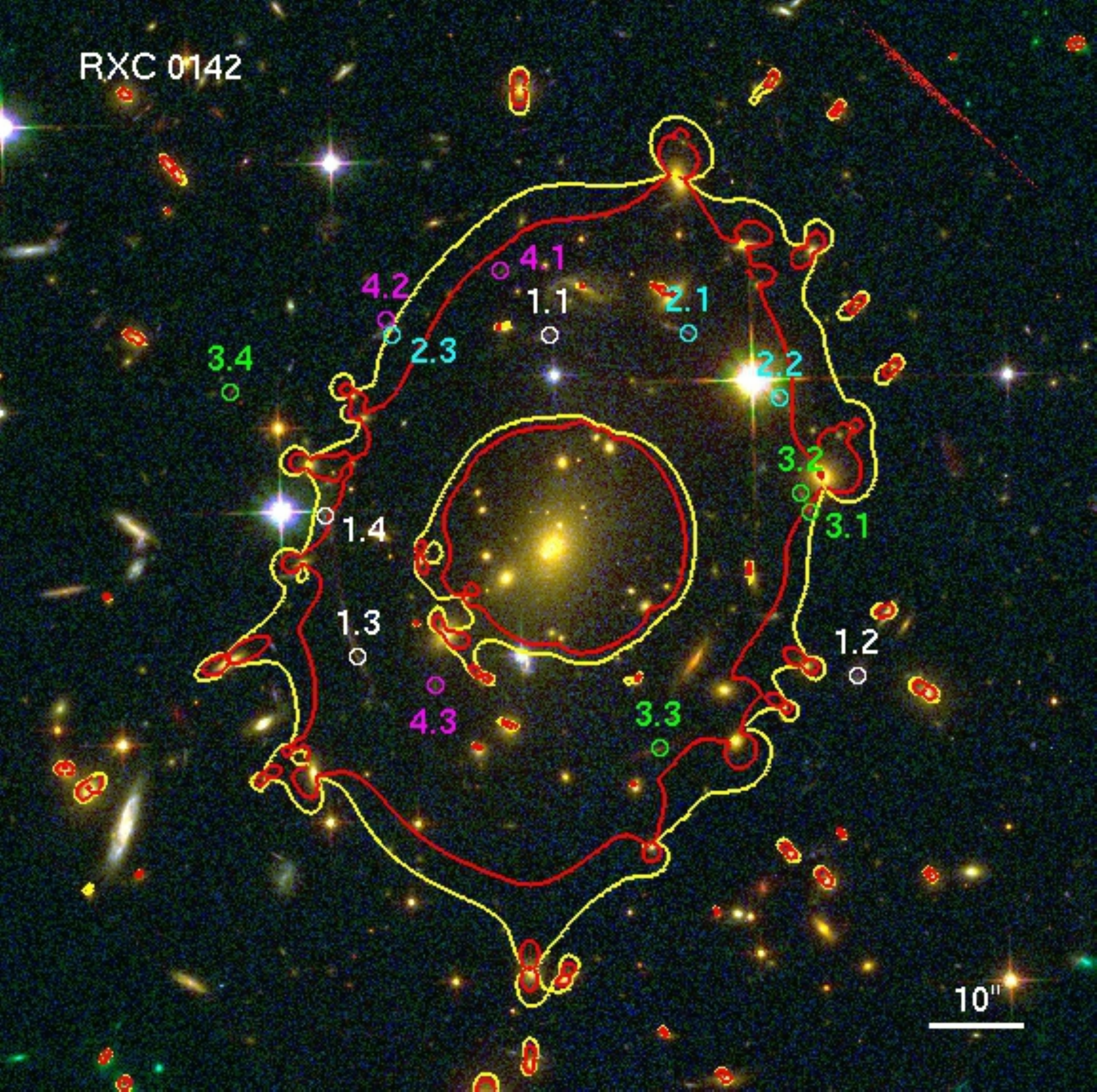}
        \hspace*{0.5cm}
        \includegraphics[width=0.49\linewidth]{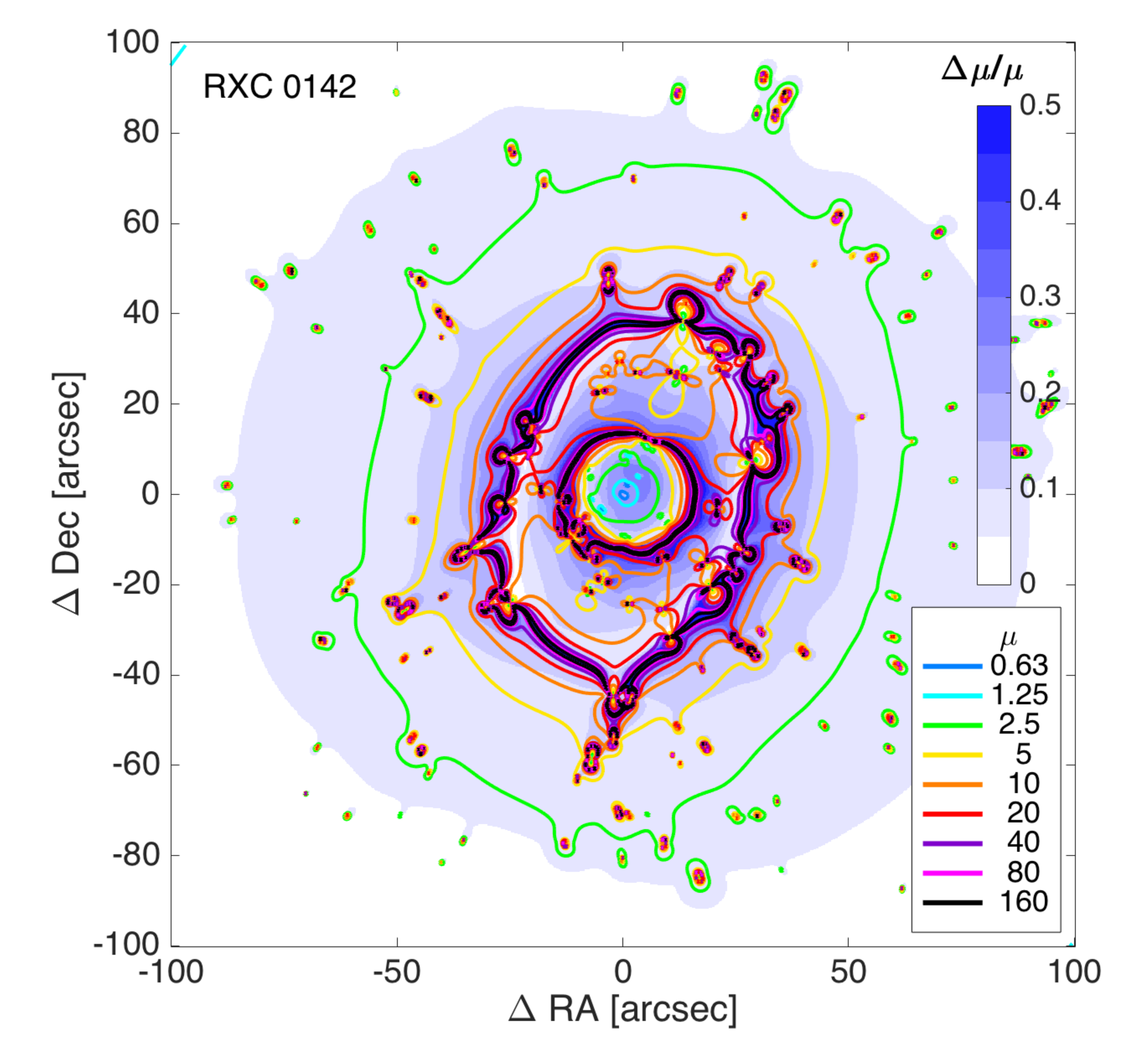}
        \caption{{\it Left}: Image of RXC~J0142+44 created using a combination of WFC3/IR imaging in the red (F160W) and ACS imaging in the green (F814W) and blue (F435W) filters. The locations of multiply imaged lensed galaxies are overplotted and labeled. The red curve marks the location of the critical curve for a source at $z=3.0$. The yellow curve marks the location of the critical curve at $z=9.0$. The image is oriented North-East. {\it Right}: The magnification for a source at $z=9.0$ from the best-fit lens model is shown as contours; contour values are given in the legend. 
        The background colormap indicates the uncertainty level in in each location, given as $\Delta\mu/\mu$. The uncertainty estimate takes into account the source redshift uncertainty, but does not account for other systematics (see text).}\label{fig.0142}
    \end{minipage}%
\end{figure*}

We identify four lensed galaxies in the field of RXC~J0142.9+4438. Their positions and  photometric redshifts are listed in Table~\ref{tab.0142arcs}. No spectroscopic redshifts are available for the cluster. 

System \#1 is labeled 1.1-1.4 in Figure~\ref{fig.0142}. The photometric redshift for this system varies significantly between the four images, due to contamination from objects near images 1.3 and 1.4. The photometry of images 1.1 and 1.2 is not affected by surrounding objects. {These images} are thus used to pinpoint a suitable photometric redshift for the system, with a photo-z around $z=1.8$. The confidence interval of this system is formally $[0.05-3.86]$; however, the range below $z=0.34$ is ruled out, since the galaxy must be behind the cluster to be lensed. 

System \#2 consists of one galaxy lensed three times, labeled 2.1-2.3. 
Due to contamination from nearby sources or low signal to noise, we were unable to measure a reliable photometric redshift for this system. We therefore set a broad prior on its redshift.   

System \#3 is also lensed four times, labeled 3.1-3.4. We fix the redshift of the system to $z=3.1$ in order to constrain the redshifts of the other systems in the field. We discuss the implication of this approach in Section~\ref{s.uncertainties}. 

System \#4 has three visible images, labeled 4.1-4.3. One additional image is predicted to be west of 3.1 and 3.2, but its 
predicted magnitude is too faint to be detected in the current data.

We find that one cluster-scale halo plus galaxy-scale halos are sufficient to reproduce the lensing observables. Table~\ref{tab.0142arcs} lists the locations of the images of the lensed galaxies and their redshifts. Table~\ref{tab.0142model} lists the best-fit parameters of the resulting model.

\subsection{RXC~J2211.7-0349}
\begin{figure*}
    \centering%
    \begin{minipage}{1.0\textwidth}
        \includegraphics[width=0.45\linewidth]{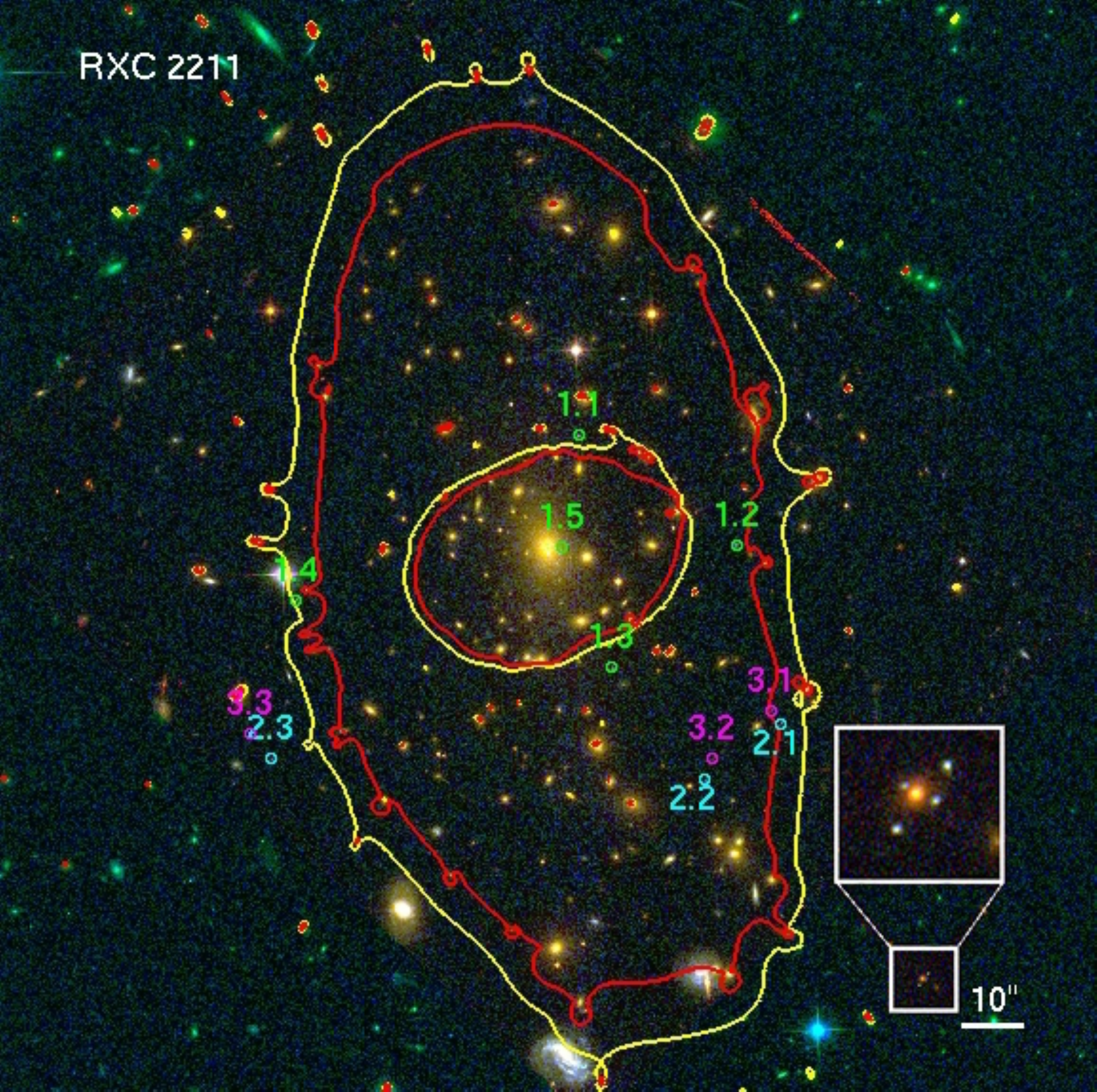}
        \hspace*{0.5cm}
        \includegraphics[width=0.49\linewidth]{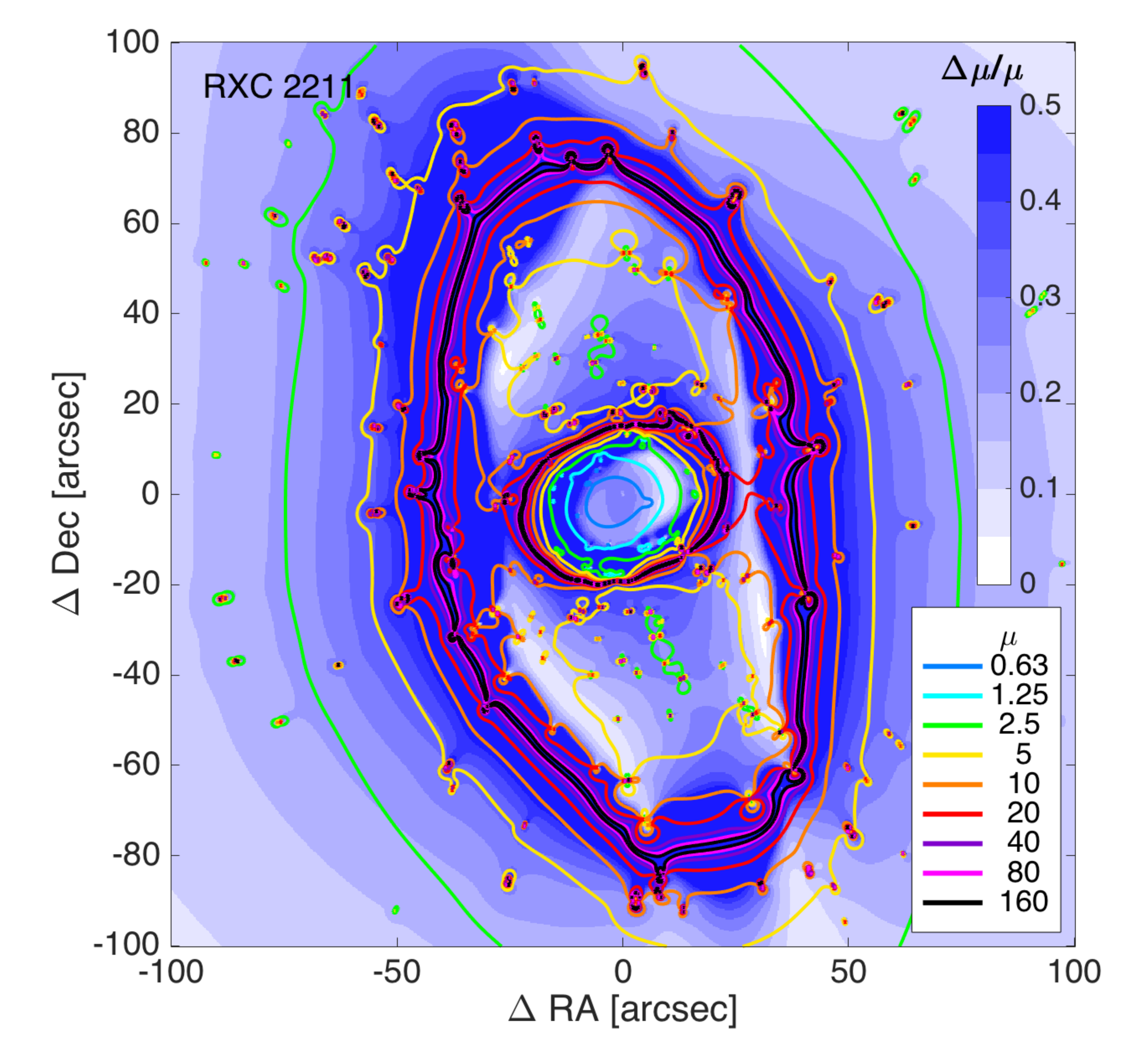}
        \caption{{\it Left}: Image of RXC~J2211.7-0349, from WFC3/IR imaging in the red (F160W) and ACS imaging in the green (F814W) and blue (F435W). The locations of multiply imaged arcs are plotted and labeled. The red curve marks the location of the critical curve for a source at $z=3.0$. The yellow curve marks the location of the critical curve at $z=9.0$. The image is oriented North-East. {\it Right}: The magnification for a source at $z=9.0$ from the best-fit lens model is shown as contours; contour values are given in the legend. 
        The background colormap indicates the uncertainty level in in each location, given as $\Delta\mu/\mu$. The uncertainty estimate takes into account the source redshift uncertainty, but does not account for other systematics (see text).}
        \label{fig.2211}
    \end{minipage}%
\end{figure*}

We detect two lensed galaxies in the field of RXC~J2211.7-0349. Their positions and photometric redshifts are listed in Table~\ref{tab.A2537arcs}.

System~\#1 is lensed by the cluster into five multiple images, labeled 1.1-1.5 in Figure~\ref{fig.2211}. {We note that the demagnified image 1.5, located close to the BCG, is clearly detected in the data, and its colors and morphology match the other images. Its blue color distinguishes it from the BCG light.} Images 1.1 and 1.2 were targeted for spectroscopy using a multi-object slit mask with LDSS3 on Magellan, which resulted in spectroscopic redshift of $z_{spec}=1.051$ based on emission lines from [OII], [NeIII] 3869, and H$_\beta$. The photometric redshifts of these images favor a lower redshift, with a 95\% CL range of $[0.821-0.979]$ where the highest likelihood is at $z_{phot}=0.85$. The spectroscopic redshift is outside the 95\% CL, but has non-zero BPZ probability for the observed $z_{spec}$. In the lensing analysis of this cluster we use the spectroscopic redshift of source \#1 as a fixed constraint.

System \#2 and system \#3 are lensed into three images labeled 2.1-2.3, 3.1-3.3, respectively. System \#3 was targeted for spectroscopy; however, we were unable to secure a spectroscopic redshift from the data. 

We identify an Einstein Cross configuration around a galaxy $90\farcs2$ from the cluster core at $R.A.$=22:11:41.986, $Decl.$=-03:50:52.31. The F606W-F814W color of the lensing galaxy is redder than the cluster red sequence, likely placing this lens galaxy behind the cluster. The lensing potential of this galaxy is boosted by the nearby lensing potential of the cluster, thus allowing it to reach a critical  mass density for strong lensing. The projected distance of this lensing galaxy is far enough from the center of the cluster that it does not significantly affect the cluster lensing potential. We therefore exclude this galaxy and lensed images from the model and leave an analysis of this lensing galaxy to future work.

We find that two cluster-scale halos are required in order to match the lensing observables, supplemented by galaxy-scale halos. Table~\ref{tab.2211arcs} lists the locations of the lensed galaxies and their redshifts. Table~\ref{tab.2211} lists the best-fit parameters of the resulting model. The projected mass density of this cluster appears to be a ``warped'' elliptical, with a radius-dependent position angle. This may indicate a more complex dark matter distribution than what may be {implied} from the galaxy distribution. A multiwavelength and dynamical analysis of this cluster may shed more light on its structure. 

\subsection{ACT-CLJ0102-49151}
\begin{figure*}
    \centering%
    \begin{minipage}{1.0\textwidth}
        \includegraphics[width=0.45\linewidth]{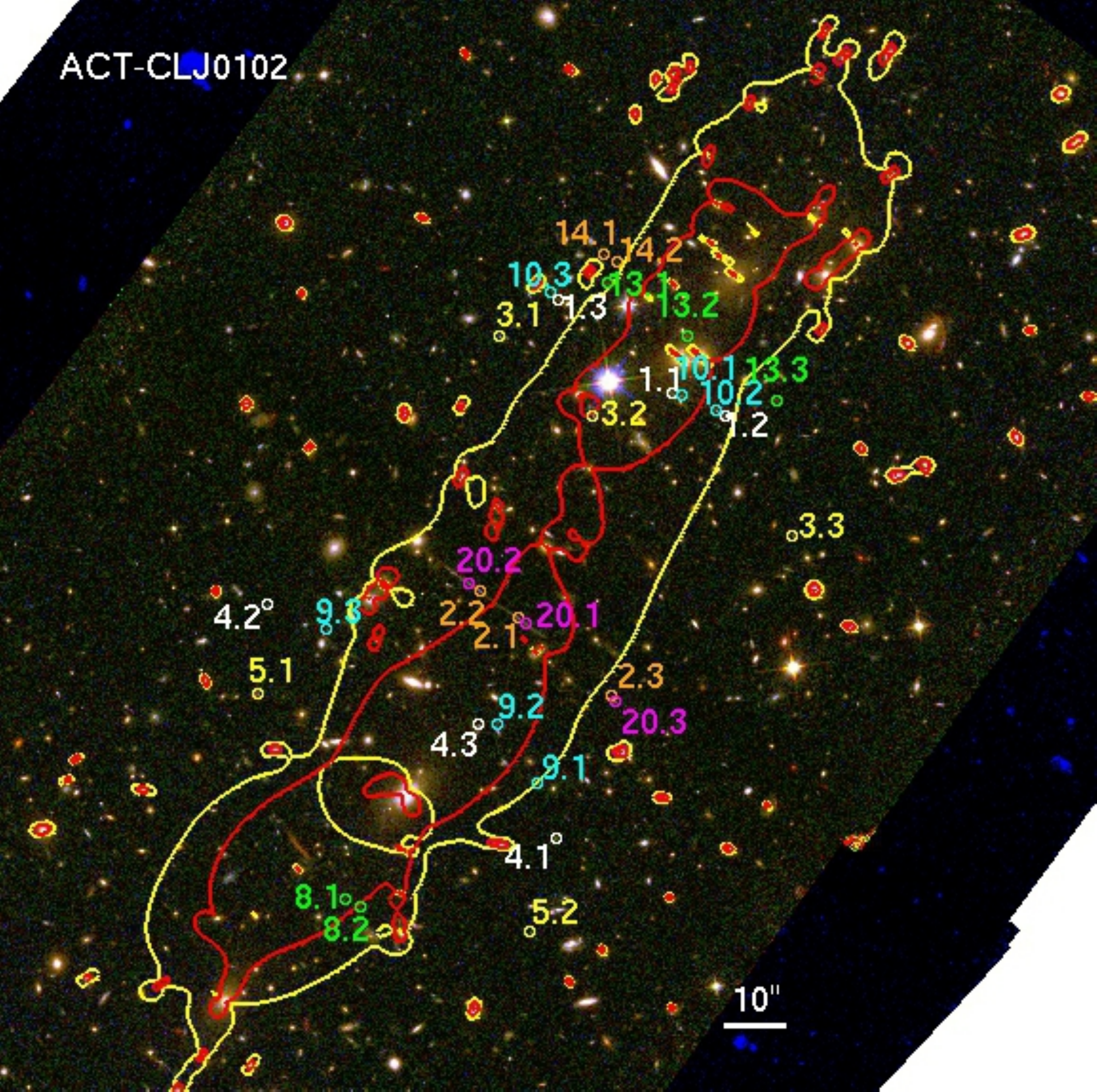}
                \hspace*{0.5cm}
       \includegraphics[width=0.49\linewidth]{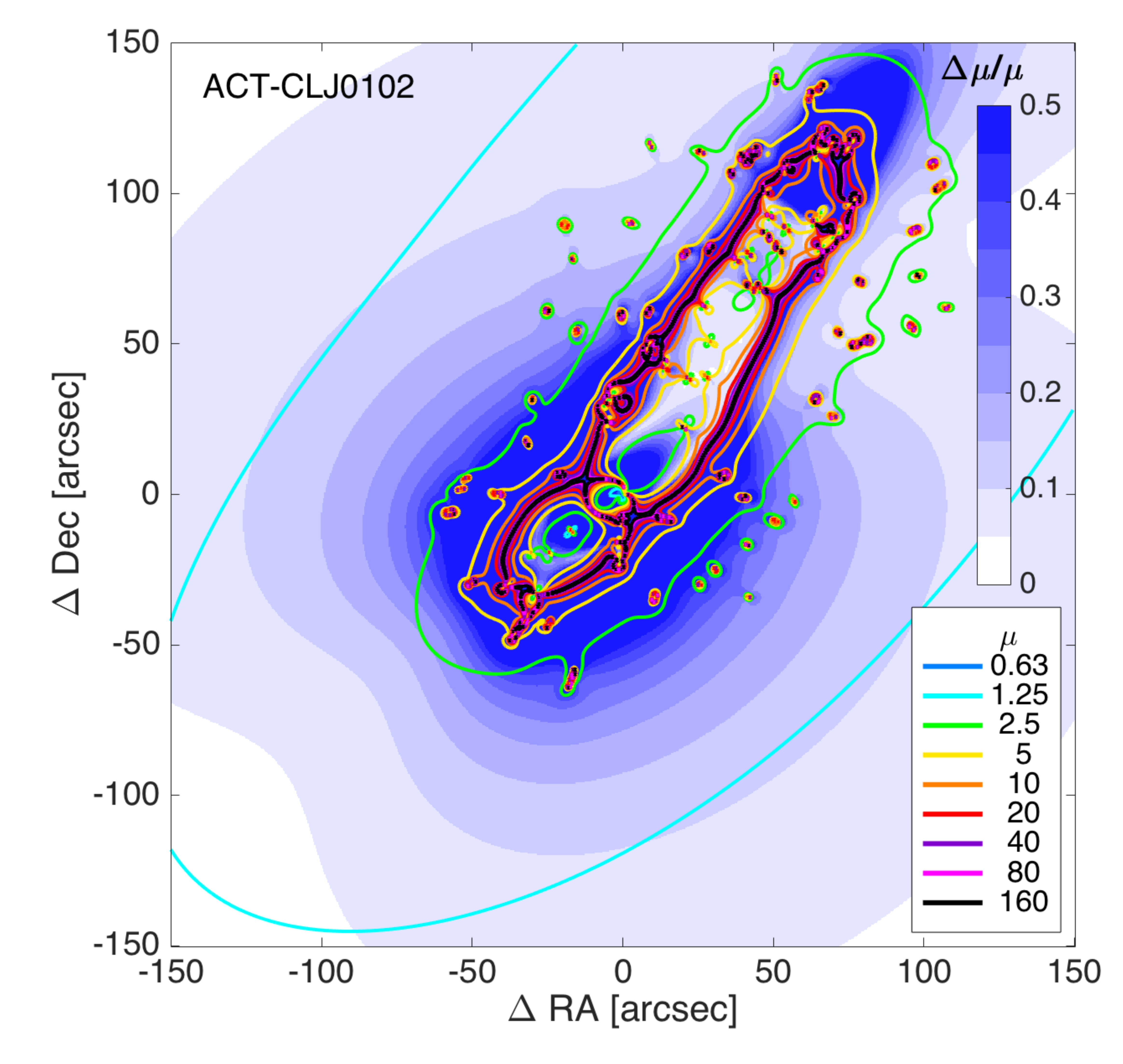}
        \caption{{\it Left}: Composite color image of ACT-CLJ0102-49151 created using a combination of WFC3IR imaging (red: F160W, green: F105W), and ACS imaging (blue: F775W). The images of lensed galaxies are overplotted and labeled. The red and yellow curves mark the location of the critical curves for a source at $z=3.0$ and $z=9.0$, respectively. The image is oriented North-East. {\it Right}: The magnification for a source at $z=9.0$ from the best-fit lens model is shown as contours; contour values are given in the legend. 
        The background colormap indicates the uncertainty level in in each location, given as $\Delta\mu/\mu$. The uncertainty estimate takes into account the source redshift uncertainty, but does not account for other systematics (see text).}\label{fig.A0102}
    \end{minipage}%
    \hspace*{1.1cm}
\end{figure*}

We identify nine lensed systems in the field of ACT-CLJ0102-49151, also known as ``El Gordo'' (Menanteau et al. 2012). Except where noted otherwise, the photometric redshifts for  the multiple images of each system are consistent with each other, as shown in Table~\ref{tab.A0102arcs}. Spectroscopy was attempted for this cluster, but no spectroscopic redshifts for any arcs were obtained.

This cluster was previously modeled 
by Zitrin et. al (2013; hereafter Z13), using ACS imaging data taken as part of GO-12755 (PI: Hughes), in F625W, F775W, and F850LP, and ground-based data. Z13 identifies nine systems and candidate lensed galaxies and uses them to create a Light-Traces-Mass model, {as well as a parametric model that uses the same scaling relations as the models in this paper}. The new \hst\ imaging data confirms five of these systems and {reveals several potential} new systems.  
Where available, we labeled our arcs with the same designation as in Z13 for consistency. These arcs are systems \#1, \#2, \#4, \#5, and \#9. 

System \#1 has two emission knots, labeled 1.1-1.3 and 10.1-10.3 in Figure~\ref{fig.A0102}. Arc system \#1 matches the identification of Z13. Arc system \#10 is a secondary emission knot in the same system, clearly resolved from \#1 in the \hst\ data. 

System \#2 appears as a straight, elongated arc, NW of the southern core of the cluster towards the northern core. The low curvature of this giant arc is indicative of significant lensing potential on both sides, as we see in the mass distribution that was derived for this cluster, as well as in other merging clusters (e.g., the ``Bullet Cluster'', Bradac et al. 2006; Clowe et al. 2006). We identify several distinct emission clumps in this arc. Their mapping indicates that the giant arc is a merging pair of two of the images of this system, bisected by the critical curve for the redshift of the arc. The third, much less distorted image of this source is clearly identified and labeled 2.3 and 20.3 in accordance with the Z13 identification.

System \#3 is a newly identified system with three images, labeled 3.1, 3.2, 3.3.

System \#4 is labeled 4.1, 4.2, 4.3. Our lensing analysis agrees with the identification of Z13 for arcs Z13-4.1, Z13-4.4, and Z13-4.5; however, we {disagree with} the identification of arcs Z13-4.2 and candidate arc Z13-c4.3 as part of the same system. The new \hst\ data shows that the colors of these arcs are much bluer than arcs 4.1-4.3, and their predicted positions are inconsistent with our lens model. 

System \#5 corresponds to arc candidate c5.3 in Z13. We disagree with the predicted counter image candidate c5.1/2 in Z13, and identify a new image, arc 5.1, based on similar parity, orientation, and colors. 

System \#8, labeled 8.1 and 8.2, is a new identification.

System \#9 has three images; 9.1-9.2 correspond to arc candidates c9.1 and c9.2 in Z13. We revise the position of 9.3 to be slightly more northwest of candidates c9.3 in Z13; otherwise the image identifications are the same. 

System \#13, labeled 13.1, 13.2, 13.3, is a new identification, made possible by its distinct IR-Visible colors. 

System \#14, labeled 14.1, 14.2, is also a new identification of a faint pair of arcs near the northern core of the cluster.

We find that two cluster-scale halos are needed in order to produce the lensing observables, similar to the previous model of Z13. These halos are located at the two regions that appear to have the densest galaxy distribution in the cluster, in the southeast and in the northwest of the field of view. {Z13 noted that the mass ratio between the SE and NW clumps is ~1.5:1, an opposite trend from the velocity dispersion measured by Menanteau et al. 2012. Similarly to Z13, we find that within the strong lensing regime (500kpc) the SE clump is somewhat more massive than the NW clump, with a projected mass density ratio of 1.19:1.} Table~\ref{tab.A0102arcs} lists the locations of the lensed galaxies and their redshifts. Table~\ref{tab.A0102model} lists the best-fit parameters of the resulting model.

\subsection{Abell~2163}
\begin{figure*}
    \centering%
    \begin{minipage}{1.0\textwidth}
        \includegraphics[width=0.45\linewidth]{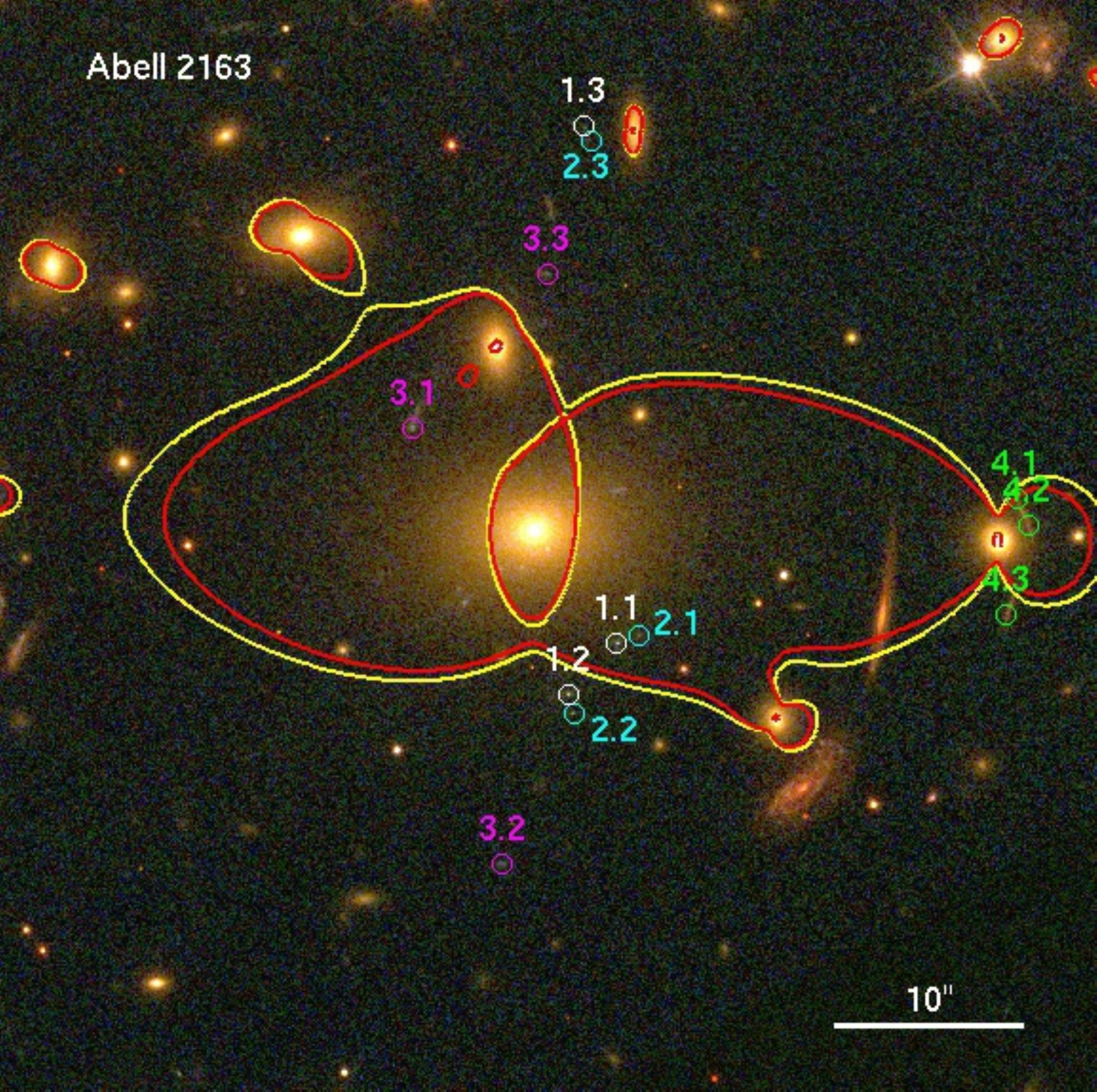}
        \hspace*{0.5cm}
        \includegraphics[width=0.49\linewidth]{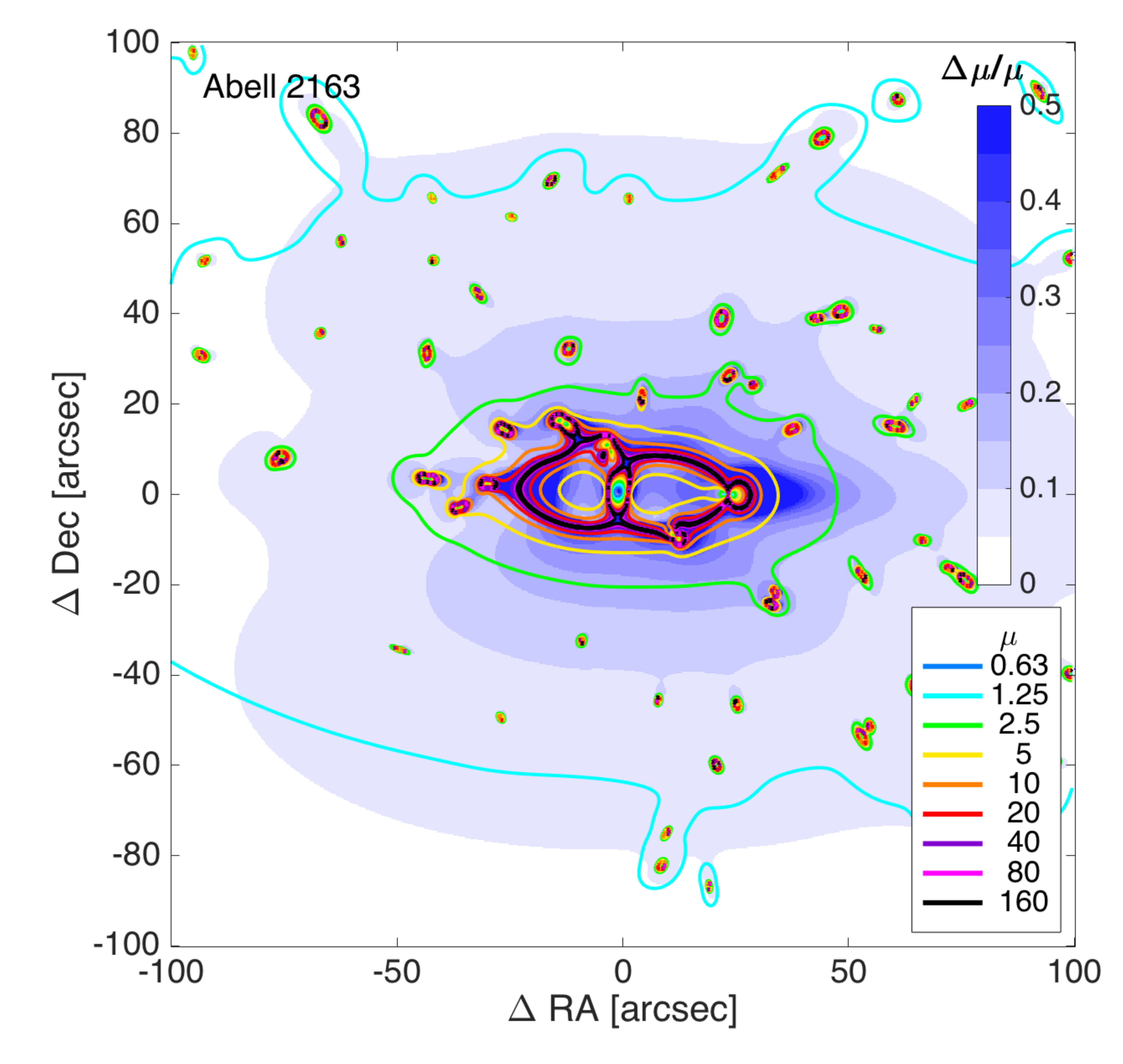}                
        \caption{{\it Left}: Composite color image of Abell~2163 created using ACS imaging (red: F814W, green: F606W, blue: F435W). The images of lensed galaxies are overplotted and labeled. The locations of the critical curves for sources at $z = 3.0$ and $z=9.0$, are marked in red and yellow, respectively. The image is oriented North-East. {\it Right}: The magnification for a source at $z=9.0$ from the best-fit lens model is shown as contours; contour values are given in the legend. 
        The background colormap indicates the uncertainty level in in each location, given as $\Delta\mu/\mu$. The uncertainty estimate takes into account the source redshift uncertainty, but does not account for other systematics (see text). }\label{fig.2163}
    \end{minipage}%
\end{figure*}


Abell~2163 is the most massive cluster in the RELICS survey with an estimated \Planck\ mass $M_{500}=16.12\times10^{14}M_\sun$. 

We identify four lensed galaxies near the core of Abell~2163, each with three images. They are labeled 1.1-1.3, 2.1-2.3, 3.1-3.3, 4.1-4.3 in Figure~\ref{fig.2163}. 

Spectroscopic observations were attempted for this cluster. However, while redshifts were obtained for several cluster member galaxies, we were unable to measure spectroscopic redshifts for any lensed galaxies used in the model. 

The redshift of arc system \#1 is fixed at $z=3.0$ based on the photometric redshift measurements of arcs 1.1 and 1.2. We note that the peak of the probability distribution function of the photometric redshift of image 1.3 is much lower, at $z=0.335$, but the redshift assumed for the system is within its 95\% CL range of [0.227-3.492]. The redshift of this system is fixed (and not left as a free parameter) in order to prevent the model from converging on $z>6$ for all the lensed systems. Such high redshifts for these lensed galaxies are unreasonable because their images are clearly detected in the F606W and F435W bands, ruling out a high redshift source.

We find that one central halo is sufficient to produce the lensing observables. Table~\ref{tab.2163arcs} lists the locations of the lensed galaxies and their redshifts. 
Table~\ref{tab.2163model} lists the best-fit parameters of the resulting model.


\section{Uncertainty Analysis}\label{s.uncertainties}

{Tables~\ref{tab.A2537model}, ~\ref{tab.0142model}, ~\ref{tab.2211}, ~\ref{tab.A0102model}, and ~\ref{tab.2163model} list the best-fit model parameters and their $1\sigma$ statistical uncertainties as derived from the MCMC sampling of the parameter space. To estimate the statistical uncertainties in the magnification and mass maps, we randomly selected 100 sets of parameters from the MCMC chain, and computed the mass and magnification of each model. The statistical uncertainty is calculated as the standard deviation of values in each pixel. As explained below, this uncertainty is likely underestimating the true uncertainty of the lens modeling process. }

{The statistical uncertainties on the magnification are typically {on the order of} a few percent in most of the field of view, except for regions close to the critical curve. However, several sources of systematic uncertainty can contribute significantly to the error budget (e.g., Zitrin et al. 2015, Johnson et al. 2016, Meneghetti et al. 2016, Rodney et al. 2016). Below, we quantify the amplitude of the systematic uncertainty that {results from} our modeling choices for the redshift parameters of sources without spectroscopic redshifts. Other sources of error include, e.g., mass projected along the line
of sight, and substructures or other correlated structures that are not
accounted for (Gruen et al. 2015; Umetsu et al. 2016; {Chiriv{\`i} et al. 2017)}.}

Three tests were conducted on each model in order to quantify the uncertainties associated with the lack of spectroscopic redshift constraints. 
As explained in Section~\ref{s.lensing}, in clusters without spectroscopic redshifts we chose one arc system with a photometric redshift close to $z=3.0$ and fixed the redshift of that system to its photometric prediction. The redshifts of the other sources were left as free parameters, with a broad range (typically  between $z_{cluster}$-8.0). We refer to this approach as ``model1'' hereafter.

The ``model2'' test left all the non-spectroscopic redshifts as free parameters, with none fixed. However, the priors on the free redshift of each system were constrained to match the photometric redshift range for each system. This range was determined by examining each photometric redshift in a system and using the lowest $z_{min}$ and highest $z_{max}$ values as conservative lower and upper boundaries. These values were drawn from the 95\% confidence interval for the photometric redshifts for each system, and thus correspond to the 2.5th and 97.5th percentile redshifts for each system.

In the ``model3'' test, the redshift of one of the systems was fixed to its most likely photo-z, as in the original model, but the priors on the free redshifts were constrained to the photometric redshift range as in model 2.

Among these three test models, model 1 has the broadest constraints on the free redshift parameters in order to account for potential photo-z outliers. We therefore use this model as the fiducial model. Comparatively, models 2 and 3 use more restrictive redshift priors. Despite the differences in the rigidity of the constraints between the three models, the final results for models 1-3 are generally consistent with each other.

We also conduct two ``extreme'' tests, which are designed to investigate the effect of fixing a model parameter to a significantly wrong redshift. 
These tests examined how the model would change if the fixed photometric redshift was shifted down to $z=2.0$ (``testE1'') and up to $z=4.0$ (``testE2''), while all other arcs remained free parameters.

These tests were not performed on RXC 2211, which has no fixed photometric redshifts. This cluster has only two lensed systems, one of which has a spectroscopic redshift. Fixing the redshift of the other system would artificially reduce the uncertainties, and we thus left {the redshift of the second system} as a free parameter. 
Abell 2537, the second cluster in this paper with spectroscopic redshift information, has a more complex model and is richer in lensing observables. The model requires two fixed redshifts to adequately constrain the strong lensing potential near its outskirts. While one of these fixed redshifts is the spectroscopic redshift acquired for system \#2, the second fixed redshift is assigned to its photo-z measurement. We therefore perform testE1 and testE2 on the fixed photo-z for system \#3 in this cluster.

{Figure~\ref{fig.modelz} serves as a diagnostic tool to rule out models that rely on false assumptions. For each model, we plot the model-predicted redshift of the sources against the photometric redshift measurements. We find that models that rely on an outlier redshift (i.e., testE1 and testE2 models) predict systematically low or high redshifts for other sources, compared to the photometric redshift probability distribution function. While we can expect a small fraction of photometric redshifts to be catastrophic failures, it is unlikely that {\it all} of them would be. We therefore argue that {this diagnostic allows us to identify extreme false redshift assumptions. Had we based a lens model on a false redshift, such as in testE1 and testE2, this model could be be ruled out as viable model based on diagnostic plots similar to those in Figure~\ref{fig.modelz}, which would force us to revise the modeling assumptions.}

For each of the test models above, we compute the lensing magnification map and compare it to that of model1. In particular, we examine the fractional error in each pixel, as $\Delta\mu = (|\mu_{test}|-|\mu_{model1}|)/|\mu_{model1}|$. From comparing model1, model2, and model3, we find that in most cases, modifying our lens modeling choice for redshift priors changes the derived magnification at the few percent level, with no significant bias. However, this deviation is in most cases significantly higher than the statistical error of either model, as derived from the MCMC sampling. This comparison indicates clearly that the MCMC statistical uncertainties underestimate the true uncertainty due to modeling choices. We therefore account for the systematic modeling error by adding it in quadrature to the statistical uncertainty (e.g., as shown in the magnification figures: Figures~\ref{fig.a2537}, \ref{fig.0142}, \ref{fig.2211}, \ref{fig.A0102}, \ref{fig.2163}).  
We note that close to the critical curve, the deviation between models increases to above the few percent level quoted above, and it is approximately at the same level as the statistical error. 

To evaluate how important these uncertainties are for studies of the background universe, we measure the extent of the field of view that is affected by high uncertainty. 
{This area} is quantified in Figure~\ref{fig.mag_unc_frac}.
As these plots indicate, our models of RXC~0142, and Abell~2163 appear to be well-constrained throughout the field of view. Comparing the magnifications of model1, model2, and model3, we find that in 90\% of the $200''\times200''$ field of view the models agree to better than 10\%. In Abell~2537, 90\% of the field of view is constrained to better than $\sim25$\%.    
The extreme error test models testE1 and testE2, which were created with significantly wrong redshift assumptions, result in significant deviation and bias. However, even if an extreme redshift error was not noticed in the modeling process, we still find that 90\% of the field is constrained to better than 20-40\%.

The {same results are not seen in the models} for ACT-CLJ0102 and RXC~2211. These two clusters appear to not have enough constraints when compared to the level of complexity of their mass distribution. The uncertainties of ACT-CLJ0102 are dominated by the statistical uncertainty, with a large range of magnification allowed by the lensing constraints. In RXC~2211, although we have spectroscopic confirmation of one of the sources, its low redshift and the fact that the small number of lensed galaxies come from only two source planes limits our ability to model this cluster with as high {a degree of} certainty as other clusters. In these two fields, the magnification uncertainty is less than 20\% in approximately 60\% of the field of view of RXC~2211, and 80\% of the field of view of ACT-CLJ0102.

\begin{figure*}
    \centering%
        \includegraphics[width=\linewidth]{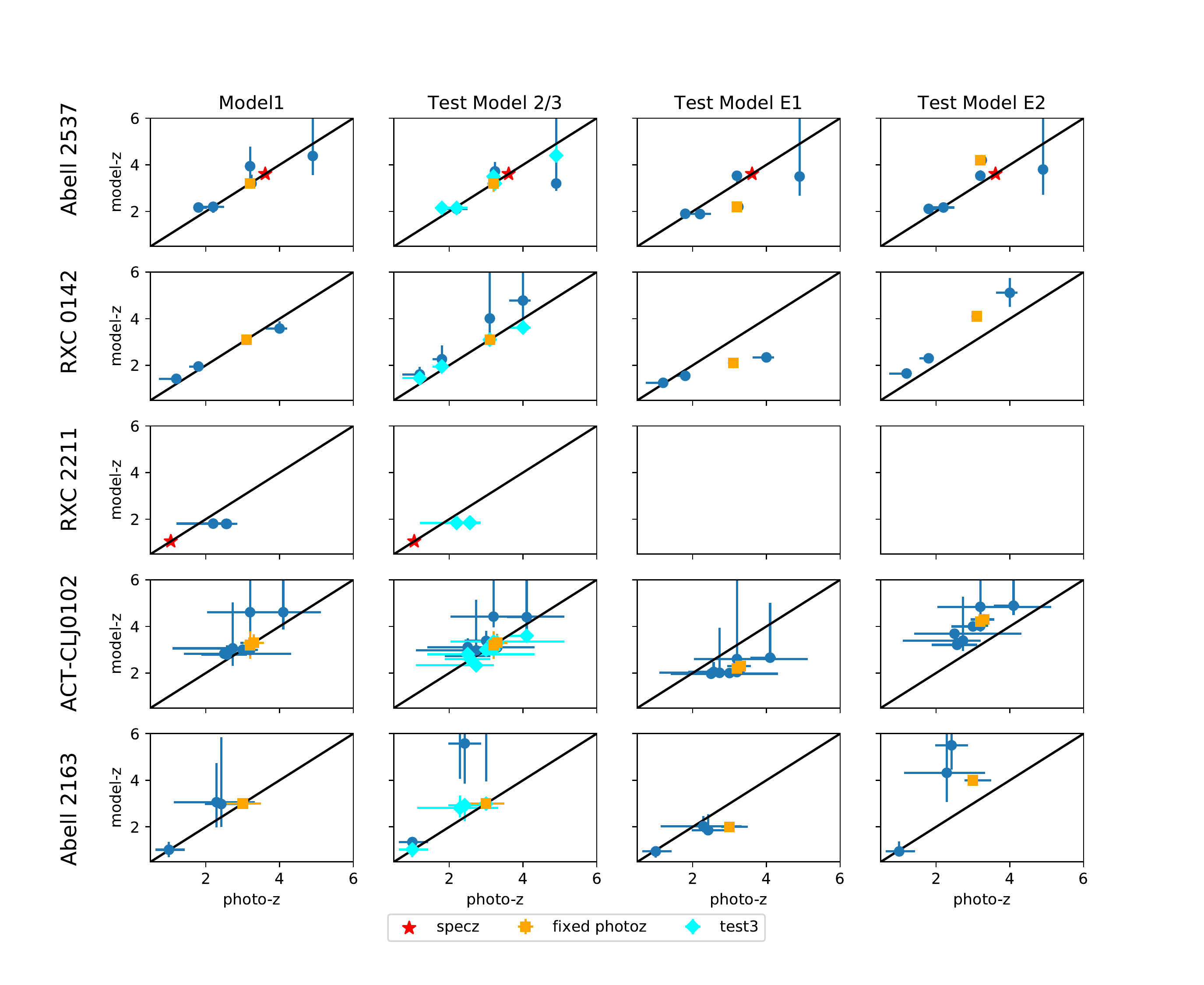}
        \caption{We plot the model-predicted redshifts against photometric redshifts for each of the test models considered in this paper. Model-z uncertainties (1$\sigma$) are from the MCMC sampling of the parameter space, and the photo-z uncertainties are from the BPZ photometric redshift analysis and represent the joint confidence limit marginalized over all the images of each system for which photo-z could be measured. Spectroscopic redshifts were always used as fixed parameters, and are labeled as `specz' in the figure. Photometric redshifts that were used as fixed parameters are labeled as `fixed'. A line with slope of unity is plotted in black to guide the eye. In tests E1 and E2, we deliberately fixed a redshift parameter to a redshift that is significantly lower or higher than its best photo-z, respectively. In these models, we find that the model systematically predicts other sources to have lower or higher redshifts when compared to their photo-zs, respectively, thus aiding in identifying whether a wrong redshift assumption was made. The two left panels show a reasonable agreement between the model-z and photo-z, confirming that the redshift assumptions that were used for these models produced reliable models, despite the small number of spectroscopic constraints. Tests E1 and E2 are not performed for RXC 2211 because its model did not use a photometric redshift as a fixed constraint.
        }\label{fig.modelz}
\end{figure*}

\begin{figure*}
    \centering%
        \includegraphics[width=\linewidth]{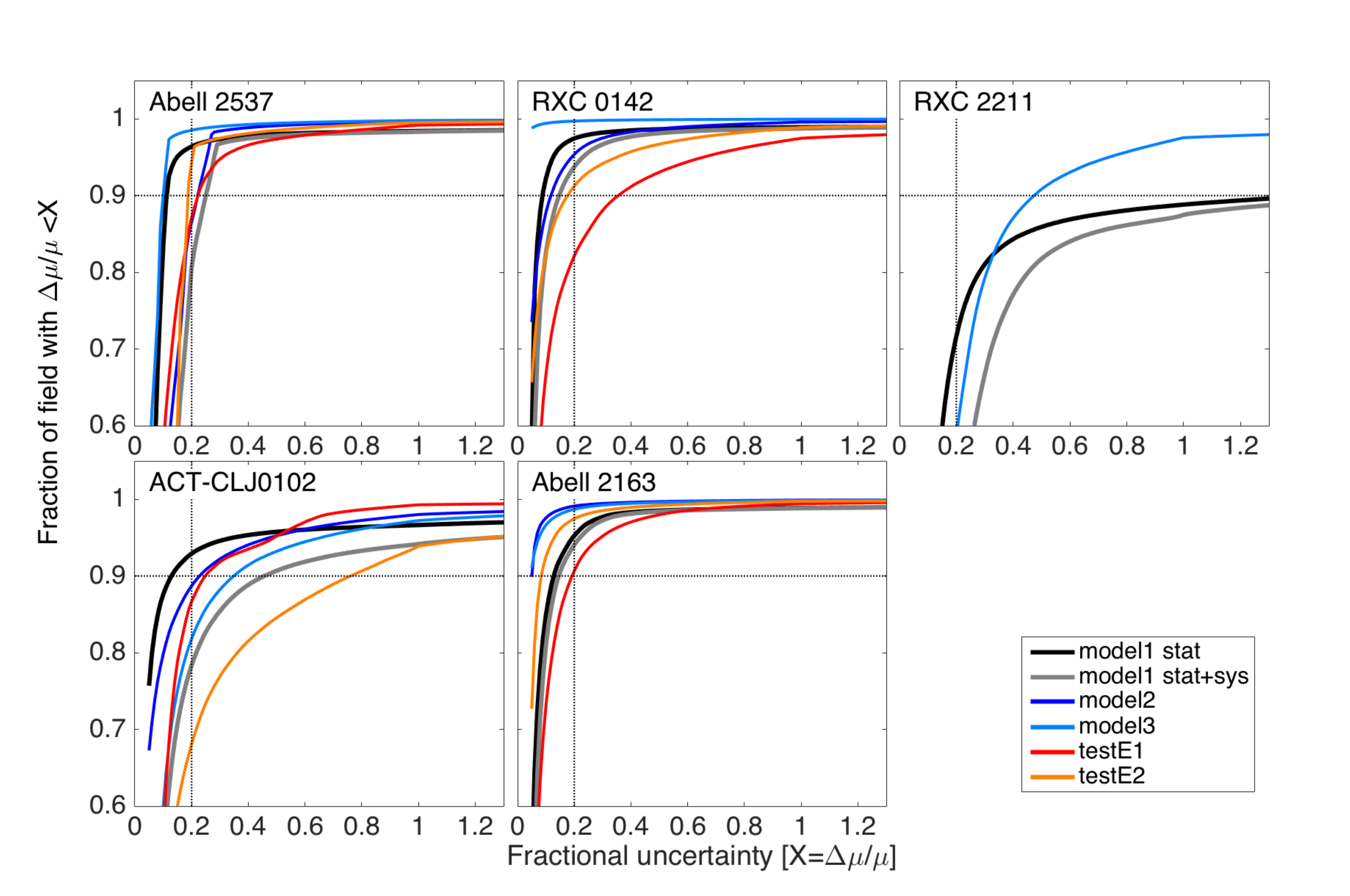}
        \caption{The fraction of the field in which the uncertainty (measured as $\Delta\mu/\mu$) is better (lower) than a given number. We show the statistical uncertainty in black, and the different uncertainty tests in colored lines as indicated in the legend. The models of testE1 and testE2 are extreme cases where the redshift is assumed to be significantly far from its best-fit value. The gray line adds in quadrature the statistical uncertainty and model2/3 uncertainty. Tests E1 and E2 are not performed for RXC 2211 because its model did not use a photometric redshift as a fixed constraint. }\label{fig.mag_unc_frac}
\end{figure*}

\section{Summary and Discussion}
\begin{deluxetable*}{lccccccc}[!bp]
\tablecolumns{8}
\tablecaption{Strong Lensing Results}
\tablehead{\colhead{Cluster }  &
            \colhead{M($<300$ kpc) }    &
            \colhead{M($<400$ kpc) }    &
            \colhead{M($<500$ kpc) }    &
            \colhead{$R_E (z=3)$}    &
            \colhead{$R_E (z=9)$}    &
            \colhead{\# sources}    &
            \colhead{\# spec-z}    \\
            \colhead{}  &
            \colhead{[$10^{14}M_\sun$]}    &
            \colhead{[$10^{14}M_\sun$]}    &
            \colhead{[$10^{14}M_\sun$]}    &
            \colhead{[arcsec]}    &
            \colhead{[arcsec]}    &
            \colhead{(\# clumps)}  &
            \colhead{}} 
\startdata
Abell~2537          &$2.0\pm0.4$&$2.6\pm0.5$ &\nodata     & $28.6 \pm 1.4$  & $32.5 \pm 1.6$ & 4 (27)   & 1   \\
RXC J0142.9+4438    &$3.4\pm0.3$&$4.5\pm0.4$ &\nodata     & $30.1 \pm 1.5$ & $33.8 \pm 1.7$ & 4 (14)   & 0     \\
RXCJ2211.7-0349     &$4.6\pm0.2$&$6.3\pm0.3$ &$7.9\pm0.4$ & $47.2 \pm 2.4$ & $52.4 \pm 2.6$ & 3 (10)   & 1    \\
ACT-CLJ0102-49151   &$5.7\pm0.5$&$8.3\pm0.6$ &$11.0\pm0.7$& $27.2 \pm 1.4$ & $40.3 \pm 2.0$ & 8 (28)   & 0   \\
Abell~2163          &$1.6\pm0.3$&\nodata     &\nodata     & $14.0 \pm 0.7$ & $14.9 \pm 0.7$ & 4 (12)   & 0    
\enddata
\tablecomments{Strong lensing analysis summary by cluster. Lensing mass is projected mass density within a projected radius of 300, 400, and 500 kpc, centered on the BCG. Errors are $1\sigma$ and include model uncertainties.  
$R_E$ is the effective Einstein radius, measured as $R_E=\sqrt{A}/\pi$, where $A$ is the area enclosed in the tangential (outer) critical curve for a source at $z=3.0$ and $z=9.0$. We list the number of unique sources, as well as the total number of multiple images of clumps that were used as constraints in parentheses.
}

\label{tab.results}
\end{deluxetable*}

{The main scientific goal of the RELICS program is to facilitate searches for high-redshift galaxies and constrain the luminosity function at the epoch of reionization. It is therefore important to not only measure the magnification due to gravitational lensing in these fields, but to also provide a good understanding of the uncertainties -- both statistical and systematic -- related to the lens modeling process. In this paper, we present strong lensing models of the first five clusters out of the RELICS program. The model outputs are available to the community through the Mikulski Archive for Space Telescopes (MAST)}.

{The accuracy of strong lens models relies on the availability of spectroscopic redshifts (e.g., Johnson \& Sharon 2016); however, only two of the clusters considered here have spectroscopic measurements of background galaxies that can be used as model constraints. Given the available resources, we anticipate that a substantial fraction of the RELICS clusters will similarly not have spectroscopic redshift constraints prior to the first JWST call for proposals. We therefore devised a strategy to appropriately handle such cases. 
In Section~\ref{s.uncertainties}, we detail our approach to minimize uncertainties due to the lack of redshift constraints, as well as our method to determine the reliability of lens models by a careful incorporation of photometric redshift information into the lens modeling process}.

At this stage, we lack spectroscopic redshifts for any multiple image set in ACT-CLJ0102-49151, Abell~2163, and RXC~J0142.9+4438. The first two clusters have been targeted for spectroscopy by the RELICS program, but these observations were unable to secure spectroscopic redshifts for any multiply-imaged sources that can constrain the models. If future observations secure spectroscopic redshifts for at least one multiply-imaged source in these fields, we will release revised lens models. Abell~2537 and RXC~J2211.7-0349 have one spectroscopic redshift constraint each, as detailed in Section~\ref{s.lensing}. 

Our models for Abell~2537, RXC~0142, and RXCJ2211.7-0349 are the first to be published for these clusters.

We revisit the previously published strong-lensing model for ACT-CL0102-49151 (Zitrin et al. 2013). The new near-infrared data from WFC3/IR provides deeper imaging data with broader wavelength coverage, which improves our ability to correctly identify multiply-imaged systems. The critical curves we derive from our model are thus slightly different than those presented in Zitrin 2013, though the overall shape of the strong lensing model is similar. 
Although our new model is based on improved image identification, the lack of spectroscopic redshifts results in large statistical and systematic uncertainties on the lensing outputs of the cluster, as seen in Figures~\ref{fig.A0102} and~\ref{fig.mag_unc_frac}. The complex mass distribution of {the cluster}, the distribution of lensing constraints, and the flexibility of the redshift parameters result in degeneracies in the parameter space. 

Spectroscopic redshifts of at least one system in each of the cluster cores of ACT-CL0102 will significantly reduce the lensing uncertainties.  
This cluster is a high mass system that is likely going through a merger (Menanteau et al. 2012), and thus has a high cross section for lensing due to the increase in both shear and convergence (e.g., Zitrin et al. 2013). 
We confirm that the elongated shape of the lens model creates an extended region of high magnification across the cluster field (Figure~\ref{fig.A0102}). 

Our work in this paper adds to previous work done for the cluster Abell~2163. Several weak lensing models have been published for this cluster (Squires et al. 1997, Cypriano et al. 2004, Radovich et al. 2008, Okabe et al. 2011, Soucail 2012). {The most direct link we can make between our strong lensing model and these weak lensing models comes from a comparison of the mass measured by weak lensing in the area covered by our model. We compare our results to the most recent of the weak lensing models, Soucail 2012, which measured a total virial mass between $8-14\times10^{14} h_{70}^{-1} M_\sun$ and provided a separate estimate of the masses of each of the sub-clumps they identified in the cluster. For the sub-clump A1, which corresponds to the area enclosed by the strong-lensing halo in this paper, they report a total projected mass density of $\sim 7.1\times10^{14} M_\sun$. In our model, we find that $\sim11\%$ of this sub-clump mass is contained within the innermost 300kpc from the BCG. This mass measurement is consistent with the assumed slope of the NFW density profile provided by Soucail 2012.}

Our strong lensing analysis of Abell~2163 is limited to the north-eastern component of this system, as we only detect multiple images that provide strong lensing constraints in this region, though large distortion is evident in individual galaxies throughout the rest of the cluster field. Our strong lens model is only well constrained at the regions where strong lensing evidence is available. 
 
We provide the effective Einstein radius for each cluster at redshifts $z=3.0$ and $z=9.0$ in Table~\ref{tab.results}. {Table~\ref{tab.results} also lists the projected mass density enclosed within radii of 300, 400, and 500 kpc from the BCG. The errors are derived from the uncertainty analysis in Section~\ref{s.uncertainties}, added in quadrature to the statistical uncertainty from the MCMC sampling. Since only one spectroscopic redshift at most is available in each field, these models are limited in their ability to reliably measure the slope of the strong lensing mass. We do not include mass density measurements for Abell~2163 at radii greater than 300 kpc because its strong lensing information is restricted to this area, and extrapolation out to larger radii will be inaccurate. Similarly, we only provide 500 kpc measurements for RXC~2211 and ACT-CL0102 because their elongated structure allows us to measure the mass density out to larger radii.

The photometric redshifts, which in RELICS are typically based on seven HST bands, provide important information to constrain the models and help reduce uncertainties. Our uncertainty analysis indicates that our overall mass measurements could have bias (mass sheet degeneracy; e.g., Schneider \& Seitz 1995) of 10\% between our test models of Abell~2163, 20\% in Abell~2537, $>$10\% in RXC~0142, and 15\% in ACT-CL0102. Unlike the other clusters, RXC~2211 has a limited number of lensing constraints. We detect lensed galaxies in two background source planes, only one of them spectroscopically confirmed. Our lensing analysis of this cluster indicates that it likely has a complex structure. With significantly fewer positional and redshift constraints compared to the other clusters, its model is less robust to changes in modeling assumptions. Both the lensing magnification and mass have higher uncertainty. The total projected mass density within a radius of $20-100\arcsec$  formally has low uncertainty of $<5$\%; however, it shows large azimuthal and mass density slope variation between test models. } 
 
{The ``lensing efficiency'' of a lens can be gauged by estimating the total magnified image-plane and source-plane area. A lens is considered more efficient when it lenses a larger area above a given magnification. Figure~\ref{fig.magplot} shows the cumulative source plane area, for a source at $z=9.0$, that is magnified above a given magnification for all the clusters considered in this paper.  
In Figure~\ref{fig.sourceplane}, we also show the effective source plane field of view that is captured by a $200''\times200''$ image in the direction of each cluster. We find that the lensing efficiency varies among clusters and depends on the cluster redshift, elongation, and concentration, and thus is not a simple function of total cluster mass.
To contextualize the results from these clusters, we compare their lensing efficiency to that of the clusters from the Frontier Fields program (shown in gray band in Figure~\ref{fig.magplot}, reproduced from Johnson et al. 2014).  
We find that most of the clusters presented in this paper have comparable or somewhat lower lensing efficiency to those of the Frontier Fields clusters. Of particular note is that despite its high mass, Abell~2163 has a significantly lower lensing efficiency than other clusters, as can be seen in Figures~\ref{fig.sourceplane} and~\ref{fig.magplot}. Its source plane is significantly less magnified, with most of the source plane area magnified by a less than a factor of 2. We speculate that this may be due to relatively low concentration. However, we leave an investigation of what causes this cluster to be a poor lens to future work. A proper analysis will benefit from a complete comparison set of lens models from the entire sample, as well as multi-wavelength measurements of the cluster's mass properties, which can be used to derive mass estimates from other mass proxies.  
}

\begin{figure*}[!hbtp]
    \centering
    \includegraphics[scale=0.3]{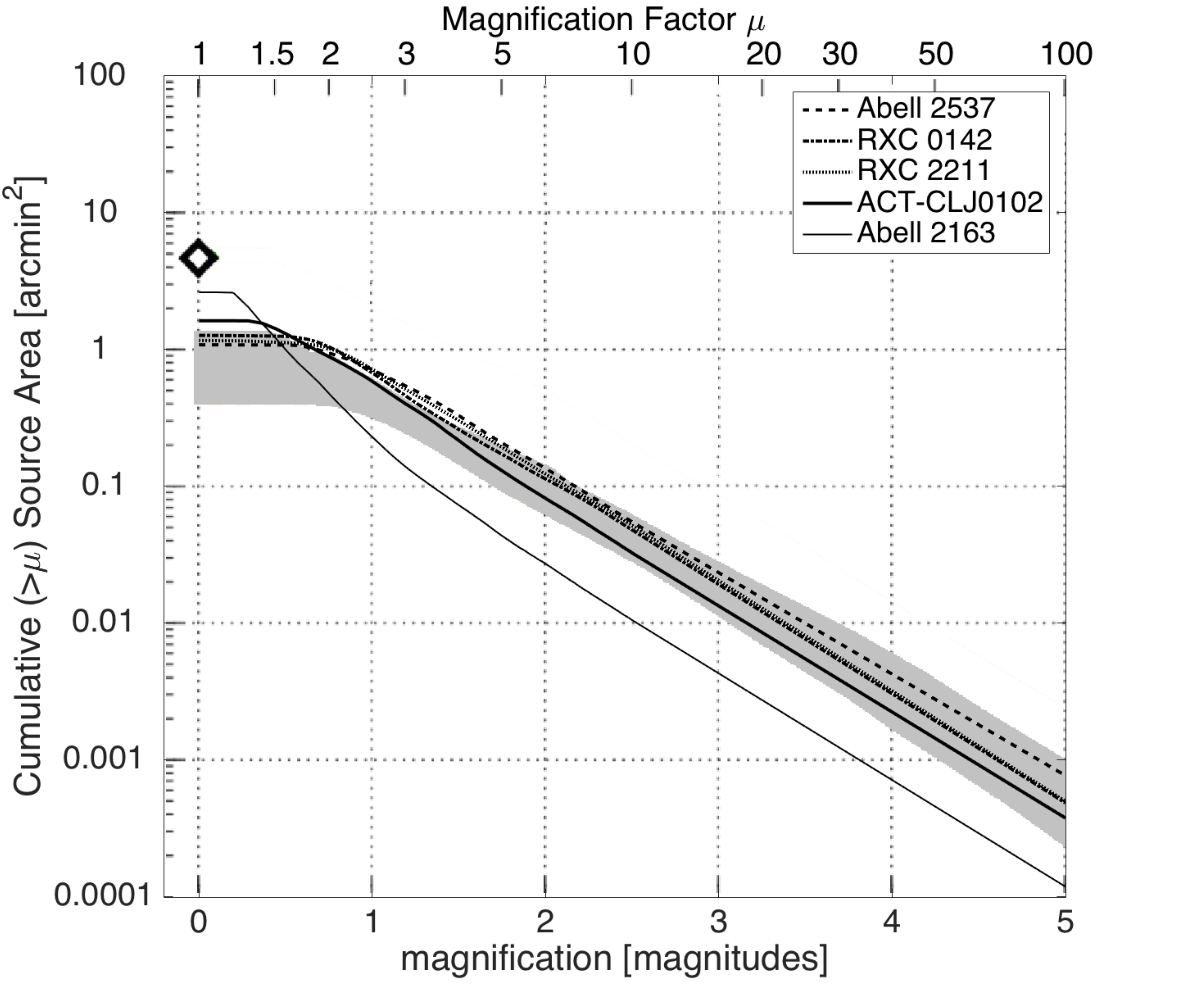}
    \caption{
    The cumulative source plane area with magnification higher than a given number, for a source at $z=9.0$. For all the clusters, we assume an image plane field of view of $130''\times130''$, which is approximately the field of view of the WFC3/IR camera. 
    We show the corresponding area of one WFC3/IR FOV with the black diamond. The shaded gray band represents the range of lensing efficiencies for the six clusters from the \hst\ Frontier Fields program (Lotz et al. 2017), reproduced from Johnson et al. (2014). We plot the lensing efficiencies for the five clusters presented in this paper on top of this band. All efficiencies are plotted for the source plane. 
    }\label{fig.magplot}
\end{figure*}

\begin{figure*}[!hbtp]
    \centering
    \includegraphics[scale=0.35]{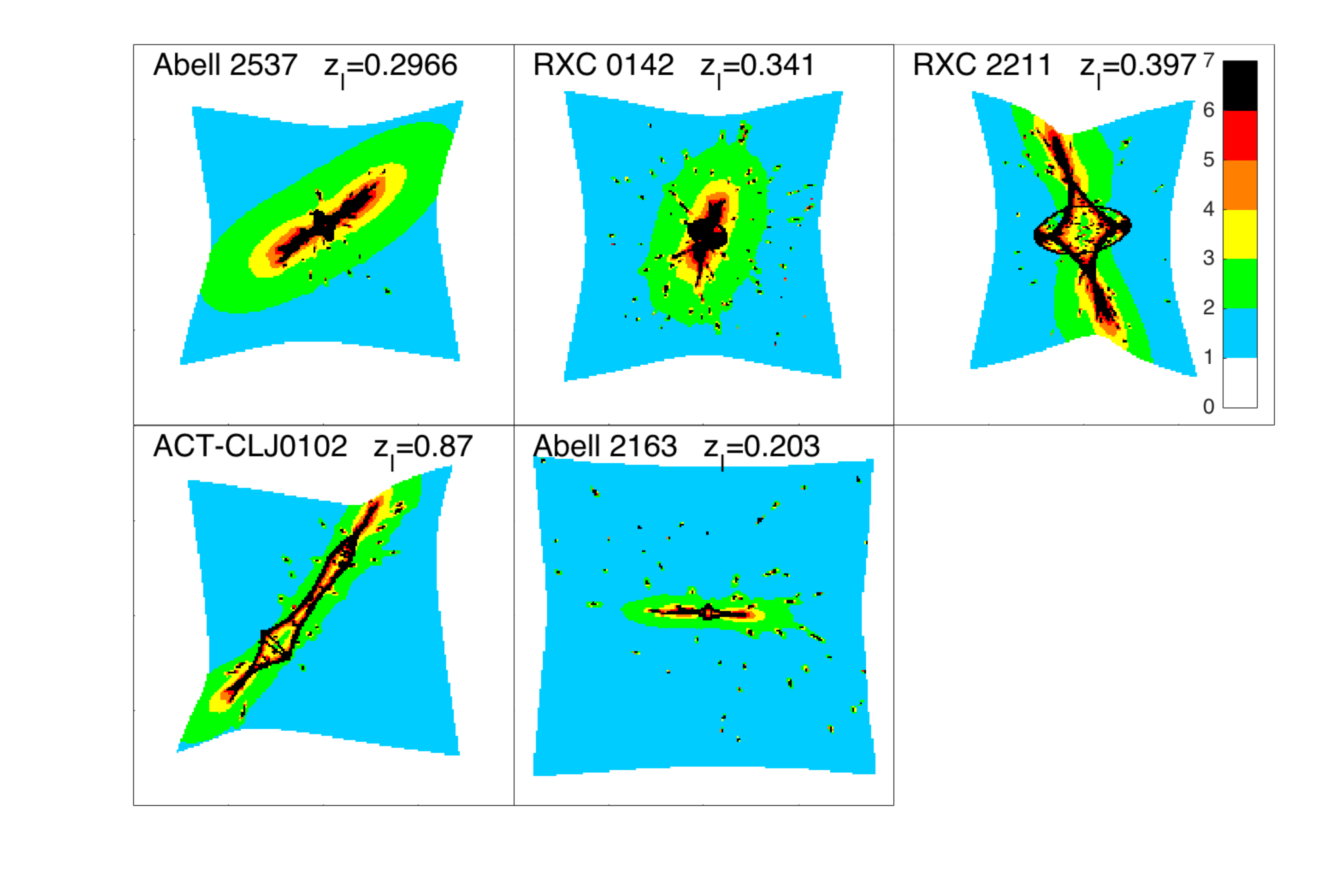}
    \caption{The $z=9.0$ source plane, color coded by magnification, to show the lensing efficiency of each cluster. Each box is $200''\times200''$ approximately the field of view of ACS. The $200''\times200''$ image plane was ray-traced to the source plane using the best-fit model1 of each cluster. Except for ACT-CL0102, all fields are centered on the BCG.}\label{fig.sourceplane}
\end{figure*}

The models presented in this paper represent a selection of the first galaxy clusters imaged by the RELICS survey. In the future, models will be computed for the remaining galaxy clusters that demonstrate strong lensing by using the same processes outlined in this paper. These models will then be used to determine the intrinsic (unlensed) properties of magnified high-redshift galaxies by the RELICS collaboration and beyond. 
High level data products from the survey, including reduced images, catalogs, and lens models, will be made publicly available through MAST. The archive will be updated with improved version of the models in this paper, as well as other RELICS fields, with the acquisition of new data to support the lensing analysis. 

\acknowledgments
This paper is based on observations made with the NASA/ESA Hubble Space Telescope, obtained at the Space Telescope Science Institute, which is operated by the Association of Universities for Research in Astronomy, Inc., under NASA contract NAS 5-26555. These observations are associated with program GO-14096. Archival data are associated with programs GO-9270, GO-12166, GO-12477, GO-12253.
Support for program GO-14096 was provided by NASA through a grant from the Space Telescope Science Institute, which is operated by the
Association of Universities for Research in Astronomy, Inc., under NASA contract NAS 5-26555.
This paper includes data gathered with the 6.5 meter Magellan Telescopes located at Las Campanas Observatory, Chile.
This work makes use of the Matlab Astronomy Package (Ofek 2014).  
F.A-S. acknowledges support from {\em Chandra} grant GO3-14131X.


\begin{deluxetable*}{lccccccc}
\tablecolumns{7}
\tablecaption{Abell~2537 -- Properties of lensed galaxies }
\tablehead{\colhead{ID }  &
            \colhead{R.A.}    &
            \colhead{Decl.}    &
            \colhead{Photo-$z$ [$z_{min}-z_{max}$]}    &
            \colhead{Spec-$z$}    &            
            \colhead{Model-$z$}    &
            \colhead{rms (")}    }
\startdata
1.1    & 23:08:21.468    & -02:11:19.17  & 0.40 [0.34--0.50]      &    -    & $2.11_{-0.17}^{+0.21}$    &  0.67  \\
1.2    & 23:08:23.244    & -02:11:35.70  & 1.83 [1.79--1.95]      &   -     &         &    \\
1.3    & 23:08:21.103    & -02:11:56.49  & 1.88 [1.74--2.03]      &   -     &         &    \\
1.4    & 23:08:23.114    & -02:11:03.16  & 1.80 [1.70--1.93]      &     -   &         &    \\
\hline
12.1    & 23:08:21.52    & -02:11:19.84  & 1.41 [0.20--3.09]      &  -      & $2.11_{-0.17}^{+0.18}$  &  0.50  \\
12.2    & 23:08:23.333   & -02:11:35.73  & 2.49 [2.20--2.65]      &   -     &         &     \\
12.3    & 23:08:21.197   & -02:11:57.50  & 0.84 [0.18--3.40]      &    -    &         &    \\
12.4    & 23:08:23.258   & -02:11:04.73  & 2.25 [1.33--2.68]      &    -    &         &    \\
\hline
2.1    & 23:08:23.738    & -02:11:23.46  & 3.95 [3.83--4.06]      &   -     & -       &  0.56  \\
2.2    & 23:08:23.494    & -02:11:10.57  & 3.81 [3.65--3.96]      &  -      & -       &    \\
2.3    & 23:08:21.802    & -02:11:16.51  & 3.82 [3.64--3.98]      &         & -       &    \\
2.4    & 23:08:22.231    & -02:11:24.58  & 3.74 [3.57--3.91]      & 3.611   & -       &    \\
\hline
20.1    & 23:08:23.681   & -02:11:14.42  & -                      &  -      & $3.58_{-0.47}^{+0.26}$   & 1.57   \\
20.2    & 23:08:23.609   & -02:11:12.72  & -                      &  -      &         &    \\
20.3    & 23:08:21.878   & -02:11:17.67  & -                      &  -      &         &    \\
20.4    & 23:08:22.162   & -02:11:22.03  & -                      &  -      &         &    \\
\hline
3.1    & 23:08:19.870    & -02:11:19.01  &  3.25 [3.19--3.31]     & -       & [3.2]   & 0.76   \\
3.2    & 23:08:19.666    & -02:11:25.42  &  3.22 [3.13--3.27]     & -       &         &    \\
3.3    & 23:08:21.132    & -02:10:54.45  &  3.24 [3.16--3.29]     & -       &         &    \\
\hline
31.1    & 23:08:19.850   & -02:11:19.93  & -                      &  -      & $3.23_{-0.35}^{+0.55}$    &  0.69  \\
31.2    & 23:08:19.694   & -02:11:24.60  &  -                     &  -      &         &    \\
31.3    & 23:08:21.276   & -02:10:53.20  & 2.69 [2.57--2.76]      &  -      &         &    \\
\hline
32.1    & 23:08:19.860   & -02:11:19.63  &  -                     & -       & $3.24_{-0.34}^{+0.45}$   &  0.65  \\
32.2    & 23:08:19.685   & -02:11:24.91  &    -                   & -       &         &    \\
32.3    & 23:08:21.235   & -02:10:53.57  &    -                   & -       &         &    \\
\hline
4.1    & 23:08:20.640    & -02:11:35.45  &  4.86 [4.76--5.03]     & -       & $4.24_{-1.68}^{+2.50}$
    & 0.69    \\
4.2    & 23:08:20.666    & -02:11:39.74  &  4.93 [4.81--5.01]     & -       &         &    
\enddata
\tablecomments{Properties of the images that were used as constraints in the lens model of Abell~2537. The model-z and rms are given for the best-fit model, and the rms is measured in the image plane for each family of multiple images.}\label{tab.A2537arcs}
\end{deluxetable*}

\begin{deluxetable*}{lccccccc}
\tablecolumns{8}
\tablecaption{Abell~2537 Model Parameters }
\tablehead{\colhead{ID}  &
            \colhead{$\Delta$R.A. (")}    &
            \colhead{$\Delta$Decl. (")}    &
            \colhead{$\epsilon$}    &
            \colhead{$\theta$ ($^\circ$)}    &
            \colhead{$r_{core}$ (")}    &
            \colhead{$r_{cut}$  (")} &
            \colhead{$\sigma$ (km $s^{-1}$)}   }
\startdata
Halo 1    & $ 10.91_{-9.20}^{+4.08} $ & $ 21.59_{-9.43}^{+8.41} $ & $ 0.68_{-0.33}^{+0.12} $ & $ 34.58_{-3.89}^{+5.58} $ & $ 83.46_{-28.74}^{+71.28} $ & $ 152.08_{-45}^{+253} $ & $ 1137_{-320}^{+513} $
\\
Halo 2    & $ 0.08_{-0.85}^{+0.61} $  &  $ 0.36_{-0.92}^{+1.49}$ & $0.24_{-0.04}^{+0.15} $ & $ 203.26_{-3.89}^{+6.36} $ &  $ 10.55_{-1.41}^{+2.00} $ & $ 451_{-63.34}^{+1.42} $ & $ 895_{-98}^{+102} $  
\\
\enddata
\tablecomments{Parameters for the best-fit lens model of Abell~2537. Error bars correspond to $1\sigma$ confidence level as inferred from the MCMC optimization. $\Delta$R.A. and $\Delta$Decl. are defined in relation to the center of the seventh-brightest cluster galaxy in the field, which is identified as the BCG, located at $R.A.$=16:15:48.948 and $Decl.$=-06:08:41.38. Position angles are measured north of west, and the ellipticity $\epsilon$ is defined as $(a^2-b^2)/(a^2+b^2)$. Square brackets indicate fixed parameters.} \label{tab.A2537model}
\end{deluxetable*}


\begin{deluxetable*}{lccccccc}
\tablecolumns{7}
\tablecaption{RXC~J0142.9+4438 -- Properties of lensed galaxies}
\tablehead{\colhead{ID }  &
            \colhead{R.A.}    &
            \colhead{Decl.}    &
            \colhead{Photo-$z$ [$z_{min}-z_{max}$]}    &
            \colhead{Spec-$z$}    &
            \colhead{Model-$z$}    &
            \colhead{rms (")}   }
\startdata
1.1    & 01:42:55.250    & +44:38:26.97  & 1.80 [1.55--1.86]     & -         & $1.97_{-0.11}^{+0.09}$   & 0.33    \\
1.2    & 01:42:52.212    & +44:37:51.41  & 1.81 [1.73--1.94]     & -         &       &   
\\
1.3    & 01:42:57.138    & +44:37:53.30  & 0.20 [0.16--0.25]     & -         &       & 
\\
1.4    & 01:42:57.460    & +44:38:08.00  & 0.11 [0.05--0.15]     & -         &       &
\\
\hline
2.1    & 01:42:53.869    & +44:38:27.13  & 1.01 [0.72--1.10]     & -         & $1.49_{-0.12}^{+0.11}$   & 0.81\\
2.2    & 01:42:52.976    & +44:38:20.35  & -                     & -         &       & 
\\
2.3    & 01:42:56.811    & +44:38:27.42  & 1.19 [1.15--1.31]     & -         &       &  
\\
\hline
3.1    & 01:42:52.684    & +44:38:08.64  & 3.11 [3.02--3.23]     & -         & [3.1] & 
1.12 \\
3.2    & 01:42:52.771    & +44:38:10.32  & 3.17 [3.06--3.24]     & -         &       & 
\\
3.3    & 01:42:54.168    & +44:37:43.53  & 3.10 [2.98--3.22]     & -         &       & 
\\
3.4    & 01:42:58.386    & +44:38:20.99  & 3.25 [0.15--3.66]     & -         &       & 
\\
\hline
4.1    & 01:42:55.725    & +44:38:33.72  & 0.44 [0.10--0.66]     & -         & $3.60_{-0.39}^{+0.71}$   & 0.33   \\
4.2    & 01:42:56.835    & +44:38:28.23  & 0.45 [0.18--3.93]     & -         &       & 
\\
4.3    & 01:42:56.864    & +44:37:48.19  & 4.05 [3.63--4.21]     & -         &       & 
\enddata
\tablecomments{ Properties of the images that were used as constraints in the lens model of RXC~J0142.9+4438. The model-z and rms are given for the best-fit model, and the rms is measured in the image plane for each family of multiple images.}\label{tab.0142arcs}
\end{deluxetable*}

\begin{deluxetable*}{lccccccc}
\tablecolumns{8}
\tablecaption{RXC~J0142.9+4438 Model Parameters }
\tablehead{\colhead{Object}  &
            \colhead{$\Delta$R.A. (")}    &
            \colhead{$\Delta$Decl. (")}    &
            \colhead{$\epsilon$}    &
            \colhead{$\theta$ ($^\circ$)}    &
            \colhead{$r_{core}$ (")}    &
            \colhead{$r_{cut}$  (")} &
            \colhead{$\sigma$ (km $s^{-1}$)}   }
\startdata
Halo 1    & $2.25_{-0.31}^{+0.22}$ & $1.10_{-0.47}^{+0.36}$  & $0.21_{-0.03}^{+0.05}$  & $70.49_{-1.18}^{+1.42}$  & $14.55_{-3.63}^{+3.36}$   & $277_{-113}^{+134}$ & $1180_{-65}^{+67}$  
\enddata
\tablecomments{Parameters for the best-fit model of the halo in RXC~J0142.9+4438. Error bars correspond to $1\sigma$ confidence level as inferred from the MCMC optimization. $\Delta$R.A. and $\Delta$Decl. are defined in relation to the center of the seventh-brightest cluster galaxy in the field, which serves as the BCG for our lens model. The BCG is located at $R.A.$ = 1:42:55.230 and $Decl.$ = +44:38:04.63. Position angles are measured north of west, and the ellipticity $\epsilon$ is defined as $(a^2-b^2)/(a^2+b^2)$. Square brackets indicate fixed parameters. $r_{cut}$ is fixed to 1500kpc for Halo 1.} \label{tab.0142model}
\end{deluxetable*}

\begin{deluxetable*}{lccccccc}
\tablecolumns{7}
\tablecaption{RXC~J2211.7-0349 -- Properties of lensed galaxies}
\tablehead{\colhead{ID }  &
            \colhead{R.A.}    &
            \colhead{Decl.}    &
            \colhead{Photo-$z$ [$z_{min}-z_{max}$]}    &
            \colhead{Spec-$z$}    &
            \colhead{Model-$z$}    &
            \colhead{rms (")}   }
\startdata
1.1    & 22:11:45.605    & -03:49:26.62  &   0.82 [0.71--0.89]    & 1.051   & -    & 0.26 \\
1.2    & 22:11:43.927    & -03:49:43.70  &   0.83 [0.73--0.94]    & -       & -    &  
\\
1.3    & 22:11:45.245    & -03:50:03.50  &   0.85 [0.71--0.94]    & -       & -    & 
\\
1.4    & 22:11:48.564    & -03:49:52.38  &   0.53 [0.42--0.63]    & -       & -    &  
\\
1.5    & 22:11:45.797    & -03:49:44.44  &   -                    & -       & -    &
\\
\hline
2.1    & 22:11:43.457    & -03:50:12.02  &  2.41 [1.96--2.60]     & -       & $1.75_{-0.11}^{+0.45}$  & 0.23    \\
2.2    & 22:11:44.256    & -03:50:21.08  &  2.59 [2.40--2.72]     & -       &       & 
\\
2.3    & 22:11:48.835    & -03:50:17.57  &  2.72 [2.43--2.85]     & -       &       & 
\\
\hline
3.1    & 22:11:43.560    & -03:50:10.02  &  2.42 [1.81--2.69]     & -       & $1.76_{-0.14}^{+0.47}$  & 0.50    \\
3.2    & 22:11:44.172    & -03:50:17.55  &  1.78 [1.69--2.08]     & -       &       & 
\\
3.3    & 22:11:49.054    & -03:50:13.57  &  2.40 [1.20--2.67]     & -       &       & 
\enddata
\tablecomments{Properties of the images that were used as constraints in the lens model of RXC~J2211.7-0349. The model-z and rms are given for the best-fit model, and the rms is measured in the image plane for each family of multiple images.} \label{tab.2211arcs}
\end{deluxetable*}

\begin{deluxetable*}{lccccccc}
\tablecolumns{7}
\tablecaption{RXC~J2211.7-0349 Model Parameters}
\tablehead{\colhead{Object}  &
            \colhead{$\Delta$R.A. (")}    &
            \colhead{$\Delta$Decl. (")}    &
            \colhead{$\epsilon$}    &
            \colhead{$\theta$ ($^\circ$)}    &
            \colhead{$r_{core}$ (")}    &
            \colhead{$r_{cut}$  (")} &
            \colhead{$\sigma$ (km $s^{-1}$)}   }
\startdata
Halo 1    & $4.76_{-8.61}^{+2.29}$  & $-4.59_{-5.40}^{+14.57}$ & $0.68_{-0.28}^{+0.12}$ & $104.01_{-4.02}^{+5.91}$  & $44.01_{-18.41}^{+11.96}$   & $192_{-42}^{+143}$   & $1348_{-201}^{+305}$  \\
\\
Halo 2    & $-2.85_{-1.34}^{+4.32}$  & $-1.80_{-0.58}^{+1.09}$ & $0.19_{-0.14}^{+0.54}$ & $70.13_{-33.92}^{+20.69}$  & $9.44_{-9.03}^{+3.13}$ & $182_{-131}^{+5}$ & $992_{-408}^{+8}$  
\enddata
\tablecomments{Parameters for the best-fit model of the halo in RXC~J2211.7-0349. Error bars correspond to $1\sigma$ confidence level as inferred from the MCMC optimization. $\Delta$R.A. and $\Delta$Decl. are defined in relation to the center of the seventh-brightest cluster galaxy in the field, which serves as the BCG for our lens model. The BCG is located at $R.A.$ = 22:11:45.928 and $Decl.$ = -3:49:44.25. Position angles are measured north of west, and the ellipticity $\epsilon$ is defined as $(a^2-b^2)/(a^2+b^2)$. Square brackets indicate fixed parameters. $r_{cut}$ is fixed to 1500kpc for Halo 1.\\
}
\label{tab.2211}
\end{deluxetable*}
\begin{deluxetable*}{lccccccc}
\tablecolumns{7}
\tablecaption{ACT-CLJ0102-49151 -- Properties of lensed galaxies}
\tablehead{\colhead{ID }  &
            \colhead{R.A.}    &
            \colhead{Decl.}    &
            \colhead{Photo-$z$ [$z_{min}-z_{max}$]}    &
            \colhead{Spec-$z$}    &
            \colhead{Model-$z$}    &
            \colhead{rms (")}    }
\startdata
1.1    & 01:02:53.478    & -49:15:16.01   & 0.41 [0.11--4.06]      &  -     & $2.99_{-0.15}^{+0.29}$  &  0.50  \\
1.2    & 01:02:52.604    & -49:15:19.68   & 0.35 [0.30--0.39]      &  -     &       &   \\
1.3    & 01:02:55.321    & -49:15:01.20   & 0.33 [0.30--3.49]      &  -     &       &     \\
\hline
10.1   & 01:02:53.328   & -49:15:16.38   & 3.39 [0.22--3.50]       & -      & [3.00] & 0.90   \\
10.2   & 01:02:52.758   & -49:15:18.72   & 2.69 [2.48--2.83]       & -      &       &    \\
10.3   & 01:02:55.389   & -49:15:00.33   & 3.13 [2.89--3.31]       & -      &       &   \\
\hline
2.1    & 01:02:55.976    & -49:15:51.25   & 3.35 [3.18--3.50]      & -      & [3.3] & 2.36   \\
2.2    & 01:02:56.571    & -49:15:47.12   & 3.37 [3.20--3.51]      & -      &       & 
\\
2.3    & 01:02:54.461    & -49:16:03.69   & 3.14 [2.89--3.28]      & -      &       &   
\\
\hline  
20.1   & 01:02:55.831   & -49:15:52.37   & 3.26 [3.07--3.54]       & -      & [3.3] & 2.13   \\
20.2   & 01:02:56.741   & -49:15:45.98   & 3.09 [2.90--3.26]       & -      &       & 
\\
20.3   & 01:02:54.403   & -49:16:04.52   & 3.44 [3.28--3.54]       & -      &       & 
\\
\hline
3.1    & 01:02:56.255    & -49:15:07.05   & 4.40 [4.27--4.47]      & -      & $7.42_{-1.72}^{+0.58}$  & 0.14    \\
3.2    & 01:02:54.741    & -49:15:19.56   & 4.21 [3.99--4.34]      & -      &       &
\\
3.3    & 01:02:51.532    & -49:15:38.47   & 4.56 [4.45--4.64]      & -      &       &   
\\
\hline
4.1    & 01:02:55.349    & -49:16:26.10   & 4.05 [3.87--4.15]      & -      & $4.61_{-0.74}^{+2.29}$  & 0.54   \\
4.2    & 01:02:59.982    & -49:15:49.54   & 4.16 [3.93--4.24]      & -      &       &    \\
4.3    & 01:02:56.593    & -49:16:08.27   & 4.05 [3.56--4.18]      & -      &       &    \\
\hline
5.1    & 01:03:00.154    & -49:16:03.37   & 2.53 [1.98--2.72]      & -      & $4.61_{-1.05}^{+3.37}$  & 0.05   \\
5.2    & 01:02:55.763    & -49:16:41.04   & 0.34 [0.27--3.36]      & -      &       &
\\
\hline
8.1    & 01:02:58.731    & -49:16:35.79   & 2.73 [0.24--3.21]      & -      & $3.06_{-0.75}^{+1.97}$  & 0.63   \\
8.2    & 01:02:58.512    & -49:16:37.03   &  1.13 [1.10--2.56]     & -      &       & 
\\
\hline
9.1    & 01:02:55.632    & -49:16:17.53   & 2.57 [2.25--2.77]      & -      & $2.79_{-0.18}^{+0.41}$  & 0.59    \\
9.2    & 01:02:56.294    & -49:16:07.72   & 2.73 [2.52--3.10]      & -      &       &    \\
9.3    & 01:02:59.057    & -49:15:53.21   & 2.35 [1.86--2.84]      & -      &       &    \\
\hline
13.1   & 01:02:54.528   & -49:14:58.65   & 2.48 [2.09--2.97]       & -      & $2.83_{-0.19}^{+0.21}$  & 0.36  \\
13.2   & 01:02:53.246   & -49:15:06.98   & 4.60 [4.20--4.99]       & -      &       &    \\
13.3   & 01:02:51.806   & -49:15:17.03   & 2.47 [2.16--2.70]       & -      &       &   
\enddata
\tablecomments{Properties of the images that were used as constraints in the lens model of ACT-CLJ0102-49151. The model-z and rms are given for the best-fit model, and the rms is measured in the image plane for each family of multiple images.}\label{tab.A0102arcs}
\end{deluxetable*}

\begin{deluxetable*}{lccccccc}
\tablecolumns{8}
\tablecaption{ACT-CLJ0102-49151 Model Parameters}
\tablehead{\colhead{Object}  &
            \colhead{$\Delta$R.A. (")}    &
            \colhead{$\Delta$Decl. (")}    &
            \colhead{$\epsilon$}    &
            \colhead{$\theta$ ($^\circ$)}    &
            \colhead{$r_{core}$ (")}    &
            \colhead{$r_{cut}$  (")} &
            \colhead{$\sigma$ (km $s^{-1}$)}   }
\startdata
Halo 1    & $48.23_{-4.74}^{+11.97}$    & $81.42_{-7.01}^{+18.12}$   & $0.65_{-0.28}^{+0.21}$    & $55.63_{-3.20}^{+1.86}$   & $19.08_{-3.97}^{+18.20}$ & $235_{-131}^{+24}$  & $1067_{-88}^{+382}$  \\
\\
Halo 2    & $-5.77_{-3.55}^{+4.48}$  & $-1.79_{-3.62}^{+4.25}$ & $0.64_{-0.16}^{+0.22}$ & $38.96_{-3.50}^{+3.02}$ & $7.06_{-7.01}^{+5.66}$ & $365_{-237}^{+24}$ & $975_{-132}^{+152}$  
\enddata
\tablecomments{Parameters for the best-fit model of the halo in ACT-CLJ0102-49151. Error bars correspond to $1\sigma$ confidence level as inferred from the MCMC optimization.  $\Delta$R.A. and $\Delta$Decl. are defined in relation to the center of the seventh-brightest cluster galaxy in the field, which serves as the BCG for our lens model. The BCG is located at $R.A.$ = 1:02:57.769 and $Decl.$ = -49:16:19.20. Position angles are measured north of west, and the ellipticity $\epsilon$ is defined as $(a^2-b^2)/(a^2+b^2)$. Square brackets indicate fixed parameters. $r_{cut}$ is fixed to 1500kpc for Halo 1.}\label{tab.A0102model}
\end{deluxetable*}


\begin{deluxetable*}{lccccccc}
\tablecolumns{7}
\tablecaption{Abell~2163 -- Properties of lensed galaxies}
\tablehead{\colhead{ID }  &
            \colhead{R.A.}    &
            \colhead{Decl.}    &
            \colhead{Photo-$z$ [$z_{min}-z_{max}$]}    &
            \colhead{Spec-$z$}    &
            \colhead{Model-$z$}    &
            \colhead{rms (")}    }
\startdata
1.1    & 16:15:48.655    & -06:08:47.34  & 3.10 [2.89--3.26]    & -         & [3.00]  & 0.10    \\
1.2    & 16:15:48.825    & -06:08:50.00  & 2.90 [2.77--3.09]    & -         &         &    
\\
1.3    & 16:15:48.773    & -06:08:20.14  & 0.34 [0.23--3.49]    & -         &         & 
\\
\hline 
2.1    & 16:15:48.578    & -06:08:46.90  &  -                   & -         & $2.87_{-0.63}^{+2.39}$  & 0.12    \\
2.2    & 16:15:48.806    & -06:08:51.06  & 2.97 [2.36--3.33]    & -         &         &    
\\
2.3    & 16:15:48.742    & -06:08:20.95  & 1.69 [1.13--2.74]    & -         &         &  
\\
\hline
3.1    & 16:15:49.372    & -06:08:36.03  & 2.36 [2.22--2.73]    & -         & $2.91_{-0.82}^{+4.33}$    & 0.21   \\
3.2    & 16:15:49.062    & -06:08:58.98  & 2.26 [1.97--2.69]    & -         &         & 
\\
3.3    & 16:15:48.900    & -06:08:27.96  & 2.65 [2.20--2.87]    & -         &         &    
\\
\hline
4.1    & 16:15:47.244    & -06:08:39.74  & 0.77 [0.45--0.92]    & -         & $1.01_{-0.42}^{+0.47}$    & 0.21   \\
4.2    & 16:15:47.203    & -06:08:41.15  & 0.53 [0.44--0.68]    & -         &         &    
\\
4.3    & 16:15:47.280    & -06:08:45.78  & 1.14 [1.09--1.24]    & -         &         &   

\enddata
\tablecomments{Properties of the images that were used as constraints in the lens model of Abell~2163. The model-z and rms are given for the best-fit model, and the rms is measured in the image plane for each family of multiple images.}\label{tab.2163arcs}
\end{deluxetable*}

\begin{deluxetable*}{lccccccc}
\tablecolumns{8}
\tablecaption{Abell~2163 Model Parameters}
\tablehead{\colhead{Object}  &
            \colhead{$\Delta$R.A. (")}    &
            \colhead{$\Delta$Decl. (")}    &
            \colhead{$\epsilon$}    &
            \colhead{$\theta$ ($^\circ$)}    &
            \colhead{$r_{core}$ (")}    &
            \colhead{$r_{cut}$  (")} &
            \colhead{$\sigma$ (km $s^{-1}$)}   }
\startdata
Halo 1    &  $1.84_{-0.97}^{+8.15}$   & $1.06_{-1.55}^{+1.05}$   &  $0.67_{-0.23}^{+0.27}$     &  $178.90_{-2.37}^{+2.21}$     &  $11.37_{-1.50}^{+15.97}$   & $587_{-347}^{+10}$ & $726_{-48}^{+267}$ \\

Halo 2    &  [-0.687]   & [0.911]   &  $0.13_{-0.13}^{+0.29}$     &  $0.38_{-0.38}^{+9.6}$     &  $0.84_{-0.50}^{+1.62}$   & $124_{-116}^{+55}$ & $321_{-56}^{+155}$ 
\enddata
\tablecomments{Parameters for the best-fit model of the halo in Abell~2163. Error bars correspond to $1\sigma$ confidence level as inferred from the MCMC optimization. $\Delta$R.A. and $\Delta$Decl. are defined in relation to the center of the seventh-brightest cluster galaxy in the field, which serves as the BCG for our lens model. The BCG is located at $R.A.$ = 16:15:48.948 and $Decl.$ = -06:08:41.38. Position angles are measured north of west, and the ellipticity $\epsilon$ is defined as $(a^2-b^2)/(a^2+b^2)$. }\label{tab.2163model}
\end{deluxetable*}

\end{document}